\documentclass{aa}  
\usepackage{natbib}
\usepackage{graphicx}
\usepackage{txfonts}
\usepackage[colorlinks=true,linkcolor=blue,citecolor=blue,urlcolor=blue]{hyperref}
\bibpunct{(}{)}{;}{a}{}{,} 

\newcommand{\mic}{$\mu$m}

\usepackage[normalem]{ulem}

\begin{document} 

  \title{The dust--star interplay in late-type galaxies at $z <$ 0.5: forecasts for the JWST}
  \author{Ciro Pappalardo\inst{1,2}, George J. Bendo\inst{3}, Mederic Boquien\inst{4}, Maarten Baes\inst{5}, Sebastien Viaene\inst{5}, Simone Bianchi\inst{6}, Jacopo Fritz\inst{7}}
  \institute{Instituto de Astrof\'{i}sica e Ci\^{e}ncias do Espa\c{c}o, Universidade de Lisboa - OAL, Tapada da Ajuda, PT1349-018 Lisboa, Portugal 
  \and
  Departamento de F\'{i}sica, Faculdade de Ci\^{e}ncias da Universidade de Lisboa, Edif\'{i}cio C8, Campo Grande, PT1749-016 Lisboa, Portugal
  \and
  UK ALMA Regional Centre Node, Jodrell Bank Centre for Astrophysics, School of Physics and Astronomy,The University of Manchester, Oxford Road, Manchester M13 9PL, UK
  \and
 Instituto de F\'{\i}sica y Astronom\'{\i}a, Universidad de Valpara\'{\i}so, Avda. Gran Breta\~{n}a 1111, Valpara\'{\i}so, Chile
  \and
  Sterrenkundig Observatorium, Universiteit Gent, Krijgslaan 281 S9, B-9000 Gent, Belgium
\and
Osservatorio Astrofisico di Arcetri - INAF, Largo E. Fermi 5, 50125 Firenze, Italy
\and
Instituto de Radioastronomia y Astrofisica, CRyA, UNAM, Campus Morelia, A.P. 3-72, CP 58089 Michoacan, Mexico
  }
  \abstract
  {In recent years, significant growth in the amount of data available to astronomers has opened up the possibility for extensive multi-wavelength approaches. In the field of galaxy evolution, such approaches have uncovered fundamental correlations, linking the dust component of a galaxy to its star formation rate (SFR). Despite these achievements, the relation between the SFR and the dust is still challenging, with uncertainties related to the physical mechanisms linking the two.}
  {In this paper, we re-examine these correlations, paying specific attention to the intrinsic properties of the dust. Our goal is to investigate the origin of the observed scatter in low-redshift galaxies, and the ability of the James Webb Space Telescope (JWST) to explore such relations in the early Universe.}
  {We defined a sample of about 800 normal star-forming galaxies with photometries in the range of 0.15 $< \lambda <$ 500 $\mu$m and analysed them with different spectral energy distribution (SED) fitting methods. With the SEDs extracted, we investigated the detection rate at different redshifts with the MId-Infrared instruments (MIRI) on board the  JWST.}
  {Dust luminosity (L$_d$) and SFR show a strong correlation, but for SFR < 2 M$_\odot$ yr$^{-1}$, the correlation scatter increases dramatically. We show that selection based on the fraction of ultraviolet (UV) emission absorbed by dust, that is, the UV extinction, greatly reduces the data dispersion. Dust masses (M$_d$) and SFR show a weaker correlation, with a larger scatter due to the interstellar radiation field (IRF) produced by stars during late evolutionary stages, which shifts the  positions of the galaxies in the dust mass--SFR plane.
  
  At $z$ = 2, more than 60\% of the galaxies in the sample are detected with F770, F1000, F1280, F1500, and F1800. At higher redshifts, the detection decreases, and only 45\% of $z$ = 8 galaxies are detected with two filters. Reproducing the expected sensitivity of the Cosmic Evolution Early Release Science Survey (CEERS) and classifying galaxies according to their SFR and stellar mass (M$_\ast$), we investigated the MIRI detection rate as a function of the physical properties of the  galaxies. Fifty percent of the objects with SFR $\sim$ 1 M$_\odot$yr$^{-1}$ at $z$ = 6 are detected with F770, which decreases to 20\% at $z$ = 8. For such galaxies, only 5\% of the subsample will be detected at 5$\sigma$ with F770 and F1000 at $z$ = 8, and only 10\% with F770, F1000, and F1280 at $z$ = 6. For galaxies with higher SFR, detection with these three filters will be possible up to $z$ = 6 in $\sim$60\% of the subsample.} 
  {The link between dust and star formation is complex, and many aspects remain to be fully understood. The scatter between SFR and dust mass, and SFR and luminosity, decreases significantly when the analysis includes dust properties. In this context, the JWST will revolutionise the field, allowing investigation of the dust--star interplay well within the epoch of reionisation.}
  \keywords{methods: data analysis - galaxies: photometry - submillimeter: galaxies}
  \titlerunning{The dust-stars interplay at $z <$ 0.5}
  \authorrunning{Pappalardo et al.}
  \maketitle

\section{Introduction}

In recent years, astronomy has definitively entered a `golden age' for scientists, with enormous growth in the amount  of data available for research. New instruments capable of extending the explored regions of the electromagnetic spectrum have favoured this breakthrough. New data motivate the community to develop new analysis methods allowing an extensive multi-wavelength approach: data released from the Sloan Digital Sky Survey (SDSS-DR16, \citealt{ahu}) provide extracted photometry and spectra in the optical bands for millions of galaxies, representing one of the largest databases with accurate redshift measurements. The UK Infrared Deep Sky Survey Large Area Survey (UKIDSS-LAS, \citealt{law}) and the 2 Microns All-Sky Survey (2MASS, \citealt{skr}) supplied data at near-infrared (NIR, between 1 and 2 $\mu$m)  wavelengths, while the Wide Field Infrared Survey (WISE, \citealt{wri}) released an all-sky survey in the mid-infrared range (MIR, between 3-22 $\mu$m). In the infrared (mid and far) we have the space telescopes {\it Spitzer}, covering the range 3.6-160 $\mu$m, and {\it Herschel} \citep{pil} with its two onboard instruments \citep[PACS and SPIRE, ][]{pog,gri}, exploring the Universe between 100 and 500 $\mu$m. Finally, we have the Atacama Large Millimeter Array (ALMA) and the NOrthern Extended Millimeter Array (NOEMA), two interferometers extending the investigation to the millimetre (mm) and submm regimes. Once collected, all these data points spanning such a wide electromagnetic range called for an increasingly detailed physical interpretation. Such interpretation can be performed by reproducing the observed spectral energy distribution (SED) with stellar and dust emission (and absorption, in the case of the latter) models. 

This last,  more technical aspect gave birth to a branch of research devoted to the problem of fitting a galaxy SED \citep{arn,bol,cid,ilb3,ocv,toj,fri2,fri3,fra,pap4,han,lej,wea}. Historically,  models reproducing the evolution of stellar populations and dust emission have been developed separately, with reliable prescriptions to build synthetic spectra of different stellar populations  on one side \citep{bru3, bre, wor,fio, lei, vaz, cha, mar,fio2}, and a thorough comprehension of dust grains physics on the other \citep{dra2, dal,tak2,tak1,zub,dra,dac2,sil,asa,calu,zhu,man,sch,pop,aoy,dev2,dev3,gin,gra,bur2,nan,del2,gal2}. The two branches finally converged, performing simultaneous fitting of both components \citep{dev, gro,dac,nol,sil2,gra}. Two of these methods, CIGALE \citep{nol,boq} and MAGPHYS \citep{dac}, are based on the assumption that the radiation produced during the star formation process by the stellar and nebular components is partially absorbed by the dust and then re-emitted in the infrared part of the spectrum, satisfying the energy conservation. New families of codes have also been produced in an attempt to combine spectral and photometric analyses, recovering, through the implementation of photoionisation codes, emission lines feature; see for example {\tt Prospect} \citep{rob}) and {\tt Prospector} \citep{lej}. Recently, even machine learning methods have been involved in this challenge, with neural network algorithms showing promising results \citep{sim}.

Ultra-violet radiation (UV) dominates in star-forming regions, making it a natural tracer of star formation. However, dust absorbs a conspicuous fraction of this emission, which is re-emitted at infrared wavelengths, indicating a complementary approach where the star formation tracer is the total infrared luminosity of a galaxy. The combined analysis of stellar and dust emission described above allows a better characterisation of the star formation process and its link with the existing dust in a galaxy. Dust luminosity correlates strongly with star formation rate (SFR), but at lower SFR the correlation becomes more scattered \citep{cal2,cle,pap2}, because for those galaxies a non-negligible fraction of dust can be heated by the radiation field produced by stars in late evolutionary stages \citep{lon, ben,ben2,ben3,via2, via}. This is a part of the problem, because dealing with real physical quantities implies the conversion of dust luminosity ---which is essentially a measure of light--- into dust mass. Dust mass is somewhat related to the SFR through the Kennicutt-Schmidt relation, which links the SFR with the amount of gas mass \citep{boq3,ben2}. This relation shows a weaker correlation because most of the dust mass resides in relatively diffuse cold grains in thermal equilibrium within the interstellar medium (ISM), in regions not harbouring star formation \citep{smi,cle,pap2,boq2}. 

The interplay between the dust and stars in galaxies is relatively complex: dust absorbs the radiation produced in star-forming regions, but it is also heated by the interstellar radiation field produced by evolved stellar populations (t > 10$^8$ yr); this cold component being the bulk of the dust mass in a galaxy. Moreover, the fraction of dust heated by young and/or old stars depends on the galaxy morphology, with old stars heating up to 90\% of the dust in early-type galaxies, decreasing to $\approx$40\% in late-type and irregular galaxies \citep{ner}. The combination of all these mechanisms is responsible for the scatter observed in the dust mass--SFR correlations. A way to overcome this problem would be to estimate the SFR of a galaxy by combining IR and UV emissions, but the lack of simultaneous data at these wavelengths severely hampers this possibility \citep{ilb}. As an example, the Herschel Astrophysical Terahertz Large Area Survey \citep[H-ATLAS]{eal}, one of the largest {\it Herschel} surveys, found reliable counterparts in UV  for only 25\% of the sources detected at 250 $\mu$m \citep{bou}. To understand the scatter in the correlation between dust and SFR in a galaxy, it is therefore mandatory to build a sample that  covers  the UV--FIR wavelength range for galaxies at higher redshifts homogenously. This approach was taken by the Dustpedia team \citep{dav3}  in the nearby Universe, who applied a multi-wavelength approach to a sample of large (optical diameter
at least 1$'$) nearby ($v <$ 3000 km s$^{-1}$)  galaxies, investigating the correlation between the dust and the interstellar radiation field produced in star-forming regions \citep[see e.g.][]{bia2,bia3,ner}.

The present paper sheds light on the mutual influence between these components at larger redshifts. To this end, we inspect the relations between the dust, the SFR, the attenuation of the UV radiation produced in young star-forming regions, and the intensity of the interstellar radiation field produced by evolved stellar populations (t > 10$^8$ yr). These are the parameters that quantify the processes cited above, and we show that by introducing them in the parameter space of the problem, the scatter between the SFR and both dust luminosity and dust mass is strongly reduced. This approach is similar to that taken by \cite{boq2}, who define hybrid SFR estimators considering the variation in the IR scaling coefficients. For this purpose, we selected star-forming galaxies lying in the main sequence with photometric coverage from UV to FIR, and we applied different SED-fitting methods to estimate relevant physical parameters. 

We also investigate how the lack of specific wavebands affects the results and their implications, showing how the result is strongly linked to the available dataset, especially at MIR and FIR wavelengths. The type of data available has profound implications for their interpretation, as it is not always possible to deal with a sample of galaxies with complete photometric coverage from UV to FIR. For those cases, the approach followed is to quantify ---using different tests--- the errors due to the lack of a particular photometric band \citep[e.g.][]{smi,cie,bua2}. We focused mainly on the MIR range and the FIR below 250 $\mu$m because, in SED fitting methods based on the so-called `energy balance' approach described above, this region corresponds to the `fulcrum' of the hypothetical lever that balances stellar and dust emission. The lack of MIR data is not a trivial problem: the bulk of WISE galaxies with optical counterparts is at $z <$ 0.5 and at lower magnitudes only a small fraction of galaxies lie at 0.5 $< z <$ 2 \citep[see Fig. 4 of][]{yan}. Despite this, even at low redshifts,  up to $\sim$ 50\% of the FIR sources of a given sample have no WISE counterpart \citep{bou}. 

A significant step forward will be realised in this field for galaxies at 2 $< z <$ 5 with the James Webb Space Telescope \citep[JWST,][]{gar}, even if this space observatory will not produce all-sky surveys. Finally, MIR and FIR data are required for this type of analysis because the wavelengths in the range of 20 $\mu$m $< \lambda <$ 400 $\mu$m can be used to determine strong constraints on the estimation of dust attenuation in SED fitting methods \citep{bua}, even if data at longer wavelengths are necessary to estimate the total dust emission.

The paper structure is as follows: in Sects. \ref{sample} and \ref{method}, we describe the sample and the analysis methods. In Sect. \ref{results} we investigate the effect of removing MIR and FIR bands, while in Sect. \ref{discussion} we analyse the origin of the scatter observed in the correlations between stellar and dust emissions. In Sect. \ref{jwst} we quantify the detection rate of the SEDs found  in Sect. \ref{method} with the Mid Infrared Instrument \citep[MIRI,][]{rie2,wri2} on board the JWST. Galaxies classified according to their physical parameters are redshifted at various epochs, with detection rates measured at different sensitivities. We present conclusions in Sect. \ref{conclusions}.

\section{Data}
\label{sample}

The construction of a proper sample with which to investigate the interplay between the dust and stellar components of galaxies is not trivial. We need MIR and/or FIR data at $\lambda <$ 250 $\mu$m to trace the dust emission, but also UV data to estimate the star formation activity not absorbed by the dust. Finally, optical and NIR bands provide information about the age of stellar populations and galaxy stellar masses. A critical point is the selection criteria: if we based our choice on FIR emission, we would select mainly dusty galaxies, biasing the analysis towards highly attenuated objects. A selection based on UV would be biased towards relatively dust-poor galaxies with weak FIR emission \citep{igl}. \cite{dun} underlined the need for large samples spanning wide ranges of dust luminosities to overcome the problem of selection effects. For this reason, we based our sample on background galaxies selected from the {\it Herschel} Virgo Cluster Survey (HeViCS, \citealt{dav2,dav1}), which provide sufficiently deep IR and submm data on a field of   $\sim$84 square degrees in size. This program observed the Virgo cluster with eight orthogonal cross scans, reaching the confusion limit at SPIRE wavelengths \citep[see][for details]{pap1}. Other surveys observed larger regions with {\it Herschel}, such as the Herschel Multi-tiered Extragalactic Survey Data \citep[HerMES, ][]{oli} and the already mentioned H-ATLAS \citep{eal}: however, these are not appropriate for our purposes. HerMES does not  homogeneously cover all the different fields that compose the survey and is more focused on high-redshift sources, biasing UV detections. \citealt{bua} found that within the approximately 2000 galaxies detected in the GOODS-Herschel survey \citep{elb}, only about 250 sources have UV and FIR data, and these are mostly at $z >$ 1. H-ATLAS covered all the field with two cross scans, which were shallower, and focused mainly on luminous infrared galaxies (LIRGs, L$_{dust} >$ 10$^{11}$ L$_\odot$), which dominate at $z >$ 0.25 \citep{smi,bou}.

Working with the background galaxies of the deep HeViCS field, \citealt{pap2} investigated a galaxy population with median L$_{dust}$ = 2.3$\pm$ 0.3 $\times$ 10$^{10}$ L$_\odot$, and stellar masses M$_\ast$ = 1.9$\pm$0.1 $\times$ 10$^{10}$ M$_\odot$, the range of values where the bulk of the galaxy population shaping the SFR density at low redshifts is found. The catalogue produced by the latter authors selects point sources detected with SPIRE at 250 $\mu$m \citep{gri} that have an optical counterpart in SDSS-DR10 \citep{ahn}. We use these sources, and complement them with ancillary data described in Table \ref{ancdata}, covering the range 0.23 $\mu$m $< \lambda <$ 500 $\mu$m, sampling both stellar and dust components. The sample has about 800 galaxies with a median redshift of $z$ = 0.1$^{0.22}_{0.7}$ (16th-84th percentile), and is 95\% complete up to $z$ = 0.6 (see \citealt{pap2} and their analysis). Galaxies are mostly star forming\footnote{In this context `star-forming' indicates galaxies that occupy neither the starburst nor the passive region in the stellar mass--SFR plane.}, populating the main sequence region on the stellar mass--SFR plane \citep{spe}. We removed galaxies hosting active galactic nuclei (AGNs) because of the high contamination in the MIR \citep{fri,hat}, adopting the criterion defined with WISE data in \cite{bon}\footnote{Adapted for WISE by \cite{jar}:

\begin{equation*}
\begin{split}
[4.6]-[12] & > 2.2\\
[4.6]-[12] & < 4.2\\
[3.4]-[4.6] & <1.7\\
[3.4]-[4.6] & > 0.1([4.6]-[12]) + 0.38
\end{split}    
\end{equation*}
}, jointly with the standard Baldwin, Phillips \& Telervich \cite[BPT, ][]{bal} classification defined in SDSS spectra.

\begin{table*}   \centering
   \caption{Data used to build the sample. Column 1: Data bands; columns 2 and 3: name and reference of the survey; column 4: wavelengths coverage; column 5: Full width at half maximum in arcseconds; column 6: sensitivity of the survey at the reference bands in AB magnitudes.}
   \begin{tabular}{ccccccc}
    \hline
    Bands & Survey & Reference & $\lambda$ coverage & FWHM & Sensitivity (5$\sigma$)\\
    (1) & (2) & (3) & (4) & (5) & (6)\\
    \hline
    \\
    FUV-NUV& GuVICS & \cite{bos} & 1539-2316 \AA & 4-5.6\arcsec & $m_{FUV}$ = 23\\
    Optical ($u, g, r, i, z$) & SDSS-DR10 & \cite{ahn} & 3000-9000 \AA & $\sim$ 1.3\arcsec & $m_r$ = 22.2\\
    NIR ($J, H, K$)& UKIDSS-LAS & \cite{law} & 1.23-2.2 $\mu$m & $\sim$ 1.2\arcsec & $m_{K_s}$ = 18.2\\
    MIR (3.4, 4.6, 12, 22 $\mu$m)& WISE & \cite{wri} & 3.4-22 $\mu$m & 6.1-12 \arcsec & $m_{W1}$ = 16.46\\
    FIR (100, 160, 250, 350, 500 $\mu$m) & {\it Herschel} & \cite{pap2} & 100-500 $\mu$m & 9-36$''$ & $m_{500}$ = 13.14\\
    \hline
   \end{tabular}\label{ancdata}
\end{table*}

\section{Model description}
\label{method}

We analysed the broad-band panchromatic dataset described in Sect. \ref{sample} with two SED fitting techniques: MAGPHYS\footnote{\tt http://www.iap.fr/magphys/} \citep{dac} and CIGALE\footnote{\tt http://cigale.lam.fr/} \citep{nol,boq}. Both methods interpret a galaxy SED as a combination of different simple stellar population libraries and dust emission models. For details about these methods and how they compare, we refer to the papers mentioned above and the analysis in \cite{pap2}, but we summarise the main features  in the following paragraphs.

MAGPHYS builds two different sets of libraries reproducing galaxy stellar and dust components: libraries containing attenuated emission from stellar populations of between 91\AA\ and 160 \mic, are built using the models of \cite{bru2} with the \cite{cha} prescription for dust attenuation, a \cite{chab} initial mass function (IMF), and exponentially decreasing SFRs with random bursts superimposed on the models. The total dust emission from MIR to submm bands is assumed to be the sum of four main components: polycyclic aromatic hydrocarbons (PAHs); a continuum emission due to small grains stochastically heated to high temperatures by the absorption of single UV photons; a component originating from the diffuse ISM emitting as a modified black body; and $\propto \kappa_\lambda B_\lambda(T)$, with the dust absorption coefficient modelled as $\kappa_\lambda \propto \lambda^{-\beta}$ and $\beta$ empirically fixed to 1.5 and 2.0 for warm and cold components, respectively \citep[see e.g.][]{dun2,mag}.

CIGALE estimates the stellar component of a galaxy, exploiting emission models that assume different IMF and stellar libraries. To be as consistent as possible with MAGPHYS parameters, in our case we adopted a \cite{chab} IMF and the SEDs of \cite{bru2} convolved with exponentially decreasing SFHs, and applied the attenuation law defined in \cite{cal}. Dust emission follows the \cite{dra} model, which assumes a composition of amorphous silicate and carbonaceous grains with an exponentially decreasing size distribution, estimated empirically from the Milky Way in \cite{wei}. The physical processes heating the dust are mainly two: (a) hot O-B stars in photodissociation regions and (b) diffuse radiation produced by a high number of stars. The intensity of the interstellar radiation field is set by $U_{min}$, while $\gamma$ represents the fraction of interstellar dust illuminated by the radiation field $U_{min}$, and takes values between 0 and 1. Other parameters introduced are the abundance of PAHs, $q_{PAH}$, which quantifies the contribution of the PAHs to the total dust emission, and the slope of the starlight intensity $\alpha$ (see Sect. \ref{sec_nowise}). 

While MAGPHYS assumes  an empirical modified black body for the dust component, CIGALE estimates the dust emission with the physically motivated \cite{dra} models, which seems to us a more appropriate choice to gain insight into the mechanisms shaping such emission. For this reason, when the analysis deals with SEDs properties, we consider only the SEDs produced by CIGALE, which are more directly linked to the physical processes occurring within the dust component. However, the results of the two methods are consistent, both in terms of physical parameters and SED shape, as shown in Fig. 5 of \cite{pap2}.

\section{Biases induced by the lack of MIR--FIR data in the SED fitting process}
\label{results}

To quantify how the lack of MIR--FIR data affects the results of SED-fitting methods, we applied the codes described in Sect. \ref{method} to the sample defined in Sect. \ref{sample}. We then repeated the analysis removing WISE or PACS data alternatively, leaving unconstrained regions in the ranges 3-60 $\mu$m and 20-250 $\mu$m, respectively. At these wavebands, the emission is mainly due to a combination of stellar, MIR continuum, and PAH emission for WISE, and warm dust in thermal equilibrium with the ISM for PACS \citep{dra}. Statistics about the quality of the fits are reported in the first row of Table \ref{res}.

Data at MIR--FIR wavelengths  are critical because both methods described in Sect. \ref{method} assume the so-called `balance technique', that is, the assumption that the energy emitted by stars of all ages and absorbed by dust is re-emitted at IR wavelengths. In this process, MIR wavelengths occupy the centre of the SEDs, with both stellar and dust emission (at least up to $\lambda <$ 6 $\mu$m), and this is crucial for the determination of some relevant physical parameters, as shown in the following sections.

\subsection{Removing MIR data}
\label{sec_nowise}

  \begin{table}
   \centering
   \caption{Median results from the different fits obtained with CIGALE. Column 1: parameter estimated with the associated standard deviation; column 2: results obtained considering all data points; column 3: results obtained removing WISE data; column 4: results obtained removing PACS data.}
   \begin{tabular}{ccccccc}
    \hline
    Parameters & All & NO WISE& NO PACS\\
    (1) & (2) & (3) & (4) \\
    \hline
    \\
    Reduced $\chi^2$ & 2.3$\pm$3.3 & 1.5$\pm$2.1 & 1.8$\pm$2.6\\
    SFR [M$_\odot$ yr$^{-1}$]   & 1.9$\pm$5 & 2.4$\pm$5 & 2.4$\pm$5\\
    M$_\ast$   [10$^{10}$ M$_\odot$] & 2.7$\pm$3 & 2.7$\pm$3 & 2.5$\pm$3\\
    L$_{dust}$ [10$^{10}$ L$_\odot$] & 2.1$\pm$5 & 2.8$\pm$6 & 2.8$\pm$6\\
    M$_{dust}$ [10$^{8}$ M$_\odot$]  & 1.1$\pm$3 & 1.1$\pm$3 & 0.7$\pm$3\\
    \hline
   \end{tabular}\label{res}
  \end{table}

\begin{figure*}\begin{center}
\includegraphics[clip=,width=.79\textwidth]{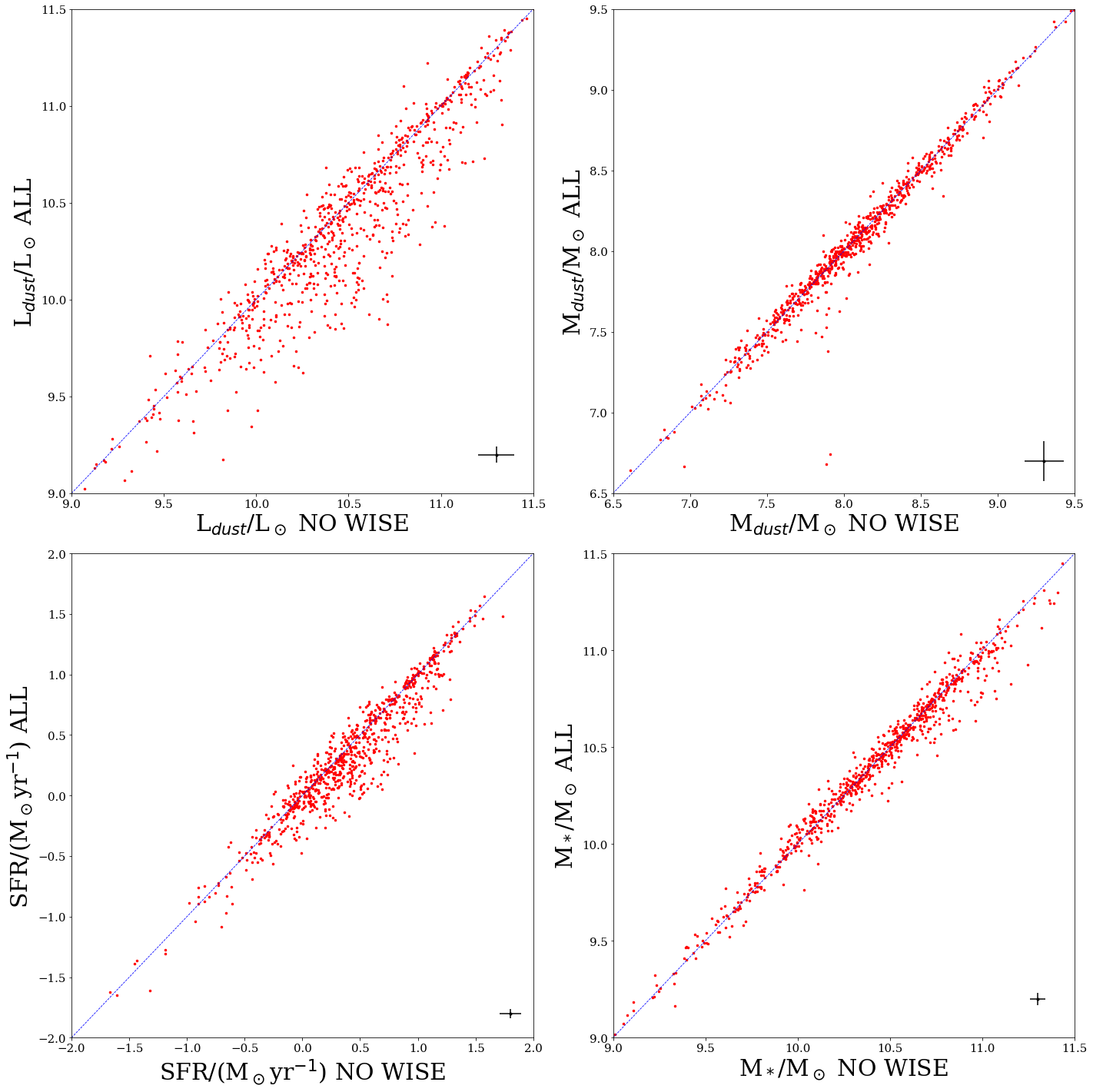}
\includegraphics[clip=,width=.99\textwidth]{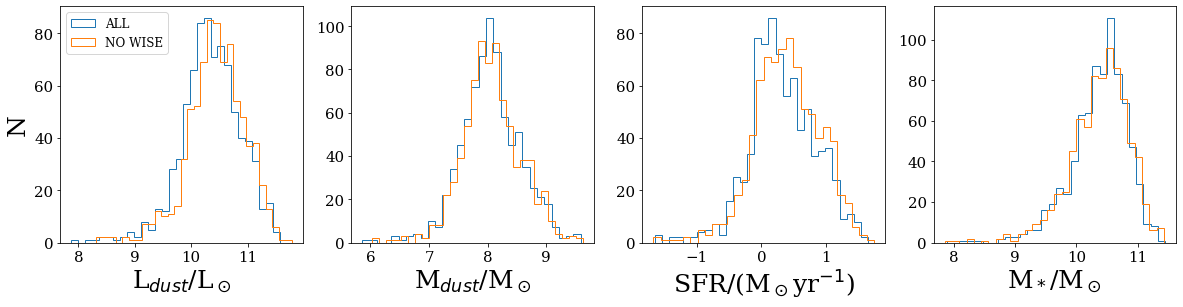}
\end{center} 
\caption{Top panels: Dust luminosity (L$_d$, top left), dust mass (M$_d$, top right), SFR (bottom left), and stellar mass (M$_\ast$, bottom right) obtained with CIGALE \citep{nol,boq} from the sample defined in Sect. \ref{sample} using the full photometric coverage (ordinate) or removing WISE data (abscissa). The blue dashed line shows a linear relation between the two quantities and the black cross in the bottom right corner shows the average errors in the fits. Bottom Panels: Histograms of the same quantities.}\label{nowise}\end{figure*}

Figure \ref{nowise} and Table \ref{res} report the differences in SFR, stellar mass (M$_\ast$), dust mass (M$_d$), and dust luminosity (L$_d$) when we exclude MIR data from the fit. The results show negligible differences in stellar and dust mass of 4\% and 7\%  respectively (right panels of Fig. \ref{nowise}), in agreement with a similar analysis by \cite{pap2}. The left panels of Fig. \ref{nowise} show SFR and dust luminosity overestimated by $\sim$28\% and 41\%, respectively. This point was also noted in a recent work by \cite{ric}, where they applied a similar approach to a set of simulated Legacy Survey of Space and Time (LSST) data to constrain physical properties of star-forming galaxies. The authors observed a clear overestimation of SFR, L$_d$, and M$_d$ when using only LSST data, highlighting for such cases the need for auxiliary rest-frame MIR observations in order to correct these overestimations. The scatters observed provided information as to how SED fitting methods based on the energy balance between the stellar and the dust component work. 

\begin{figure}\begin{center}
\includegraphics[clip=,width=.49\textwidth]{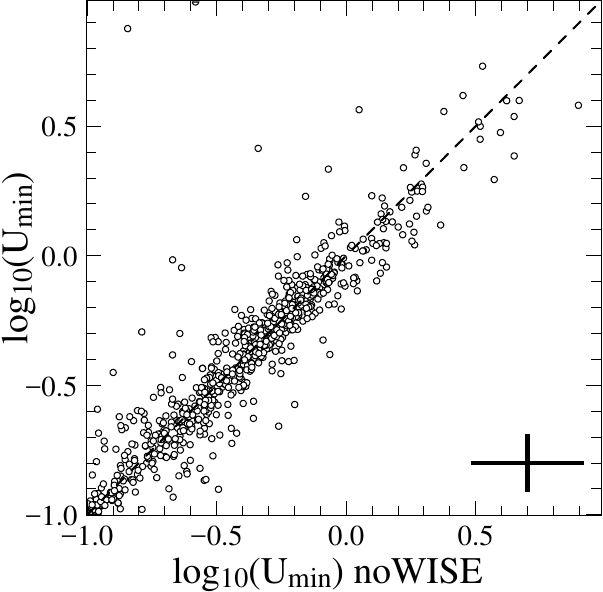}
\includegraphics[clip=,width=.49\textwidth]{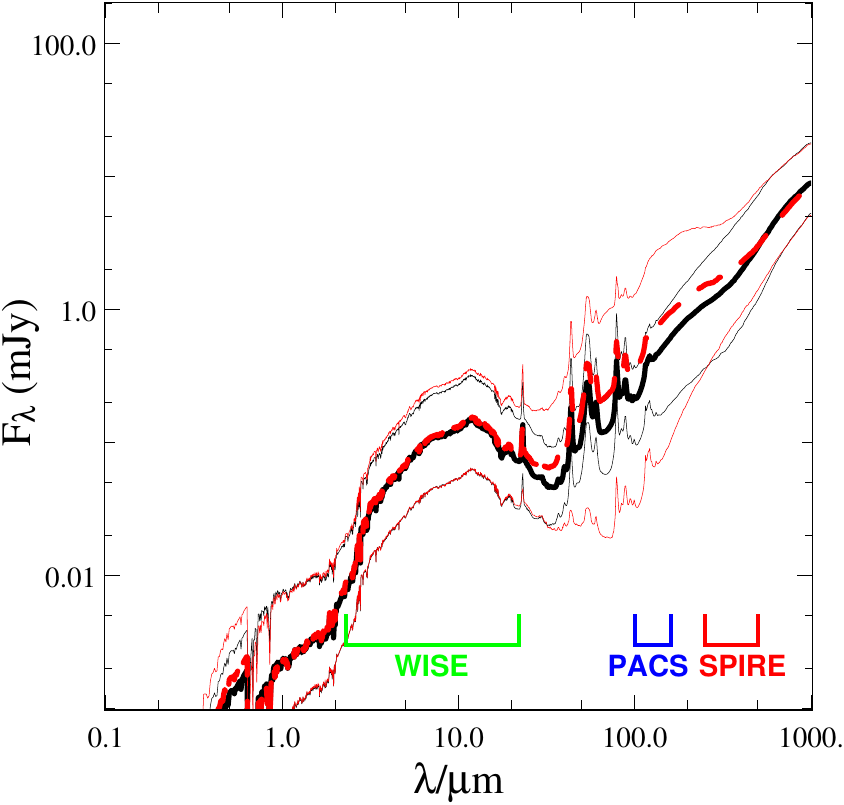}
\end{center} 
\caption{Top panel: $U_{min}$ parameter of \cite{dra} obtained with CIGALE \citep{nol} using the full photometric coverage (ordinate) or removing WISE data (abscissa). Dashed lines show a linear correlation between the two quantities and the black cross in the bottom right corner shows the average errors in the fit. Bottom panel: Median SED with 16th and 84th percentiles (ticker black solid line) obtained using the full photometric coverage (black solid) or removing WISE data (red dashed line). The photometric coverage of WISE, PACS, and SPIRE bands is shown at the bottom.}\label{umin}\end{figure}

The top right panel and the middle left histogram of Fig. \ref{nowise} show that the lack of data in the range 3-60 $\mu$m does not affect the dust mass estimation, implying that dust masses do not require MIR data in order to be reliable. This is understandable, as the emission at those wavelengths is dominated by small grains at high temperature, which despite representing a low fraction of the total dust are highly luminous. In their model, \cite{dra} parametrise this dust component through $U$, a dimensionless scaling factor that quantifies the interstellar radiation field available to heat the dust grains. CIGALE estimates a parameter, $U_{min}$, setting the typical $U$ affecting the dust. The bulk of the dust mass lies on large \citep[sub-micron sized,][]{li},
 relatively cold grains  in thermal equilibrium with the interstellar radiation field, and the consistency of the results for the dust mass with or without MIR implies that the determination of $U_{min}$ is almost unaffected by this wavelength range. Figure \ref{umin} compares the $U_{min}$ estimated using the two data sets and confirms these arguments. The average differences for the two samples are below 1\% with no dependence on $U_{min}$ values. The median SEDs obtained from the fits and shown in the bottom panel of Fig. \ref{umin} indicate that the main differences between the median spectra of the two data sets occur between 10 $\mu$m and 100 $\mu$m. At these wavebands, the primary source of dust heating is the radiation produced in the star-forming regions, and the emission is dominated by the PAHs jointly with small dust grains out of the equilibrium.

The PAH emission can be dominant at these spectral ranges (e.g. Fig. 12 of \citealt{dra}), and in Fig. \ref{umin} the median SEDs obtained without WISE bands tend to be larger than the ones recovered with the full photometric coverage. \cite{dra} quantifies the fraction of the total dust mass in the form of PAHs through the parameter $q_{PAH}$. Its variation affects mainly the spectral regions covered by the WISE data between 1 and 30 $\mu$m: the 12 $\mu$m band in particular covers several PAH features and is therefore probably the most critical for constraining $q_{PAH}$, while the WISE 22 $\mu$m band is less affected. The PAHs emission and the underlying MIR continuum are less constrained once MIR data are excluded from the fit, as also shown by \cite{ani}.

\begin{figure*}\begin{center}
 \includegraphics[clip=,width=.35\textwidth]{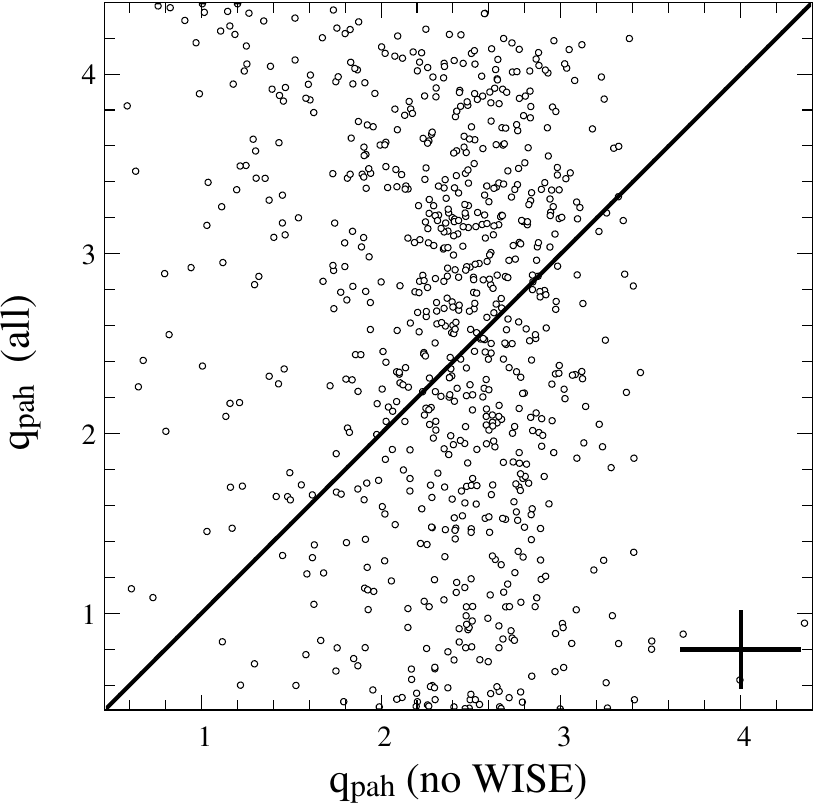}
  \includegraphics[clip=,width=.35\textwidth]{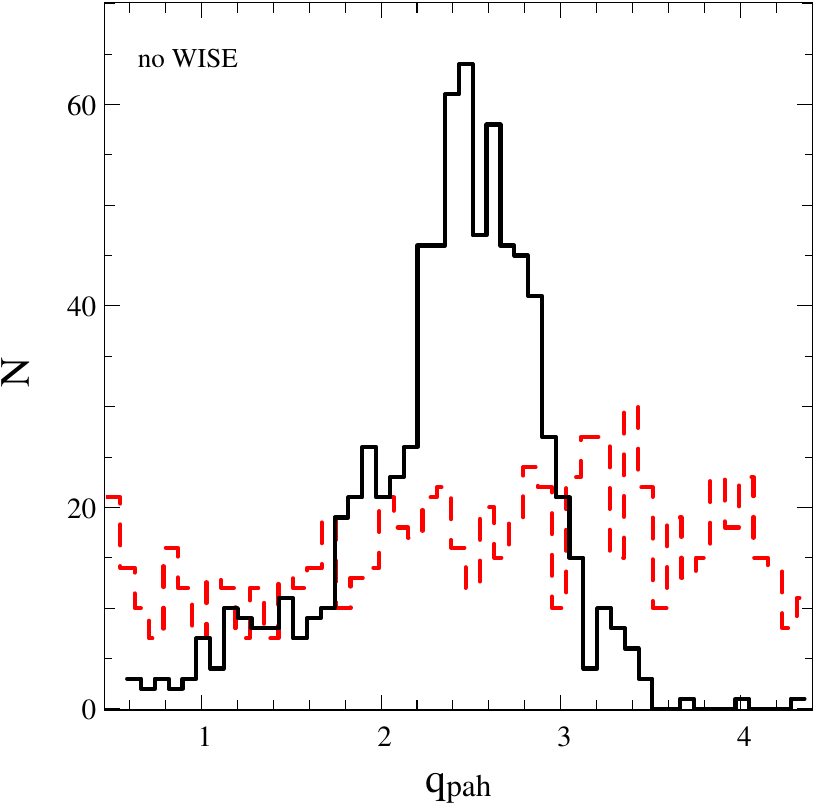}
 \end{center} 
 \caption{Left panel: $q_{pah}$ parameter of \cite{dra} obtained with CIGALE \citep{nol} using the full photometric coverage (ordinate) or removing WISE data (abscissa). Solid lines show a linear correlation between the two quantities and the black cross in the bottom right corner shows the average errors in the fit. Right panel: Histograms of $q_{pah}$ from the full photometric coverage (red dashed line) or removing WISE data (black solid line).}\label{qpah}\end{figure*}

Both components are relevant contributors to the measured dust luminosity, but the origin of the overestimation observed in the top left panel of Fig. \ref{nowise} is not due to an unconstrained fraction of PAHs. Compared to the full photometric coverage, the lack of MIR data does not lead to overestimation of PAH emission in the fitting process 
in all cases. Figure \ref{qpah} shows the $q_{PAH}$ parameter estimated with the full data set and removing WISE. The average value for both data is similar, $q_{PAH} \sim$2.4\%, but the scatter when we exclude WISE bands reduces by $\sim$50\%, as the $q_{PAH}$ values estimated in the fitting process without WISE converge around the mean of the values of $q_{PAH} \sim$2.5\%. This occurs because CIGALE tends to converge to some mean value of the grid when some parameters are not constrained. In those cases, the differences seen are somewhat grid-dependent and therefore the quantity is not constrainable.

\begin{figure}\begin{center}
 \includegraphics[clip=,width=.45\textwidth]{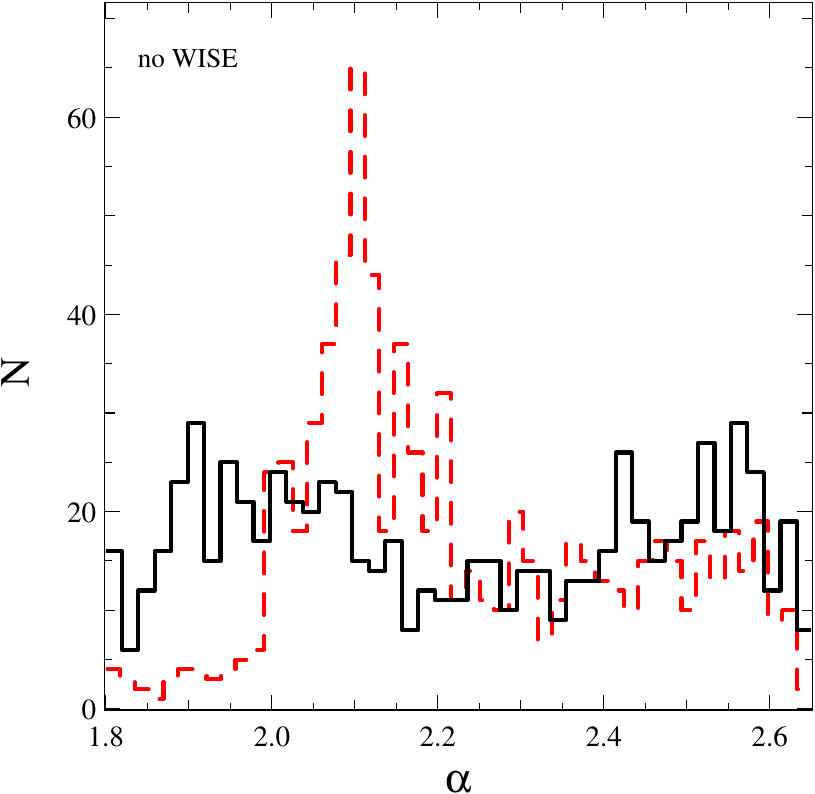}
 \end{center} 
 \caption{Histograms of the $\alpha$ parameter of \cite{dra} obtained with CIGALE \citep{nol} using the full photometric coverage (red dashed line) or removing WISE data (black solid line).}\label{alpha}\end{figure}

The real cause of the overestimation of dust luminosities  seen in Fig. \ref{nowise} is the radiation field produced in star-forming regions, the main driver of the dust heating. According to the \cite{dra} model, and similarly to \cite{dal} and \cite{dal2}, the parameter $\gamma$ quantifies the fraction of the dust mass that is exposed to starlight intensity $U_{min}$, which depends on the shape of starlight intensity. Usually, in a galaxy, this parameter follows a power law with spectral index $\alpha$. If $\alpha$ = 2, the fraction of dust mass exposed to the interstellar radiation field is equal to the fraction of the dust mass heated in star-forming clouds. A lower value of $\alpha$ implies a higher fraction of dust heated by star formation, with higher dust luminosities, while a value $\alpha >$ 2 indicates a higher fraction of cold dust overall. Figure \ref{alpha} compares the $\alpha$ values estimated using the full data set (red dashed line), and the ones obtained when removing WISE data (black line). The average value for the full data set is $\alpha$ = 2.23$\pm$0.2, which is consistent with the selection criteria adopted to define the sample in Sect. \ref{sample}. The objects analysed are moderately star-forming galaxies settled onto the main sequence; for these objects, the dust heating is powered by star formation, but also by the interstellar radiation field, as seen in the nearby Universe (see also \citealt{lon,ben,ben2,del,ben3,via2,via,boq}). When we exclude WISE data, the distribution of $\alpha$ becomes more homogeneous, increasing the scatter by a factor of two. This effect leaves the distribution rather unaffected for $\alpha >$ 2.3 but changes the distribution at $\alpha <$ 2.0  substantially, where a consistent number of galaxies have, without WISE data, a higher fraction of dust exposed to the star formation because of the modified slope of starlight intensity. For datasets lacking MIR data, this effect implies that even if the dust mass estimation is correct, the estimated fraction of dust heated in star-forming clouds is higher, leading to overestimation of dust luminosities by $\sim30\%$.

\begin{figure}\begin{center}
 \includegraphics[clip=,width=.48\textwidth]{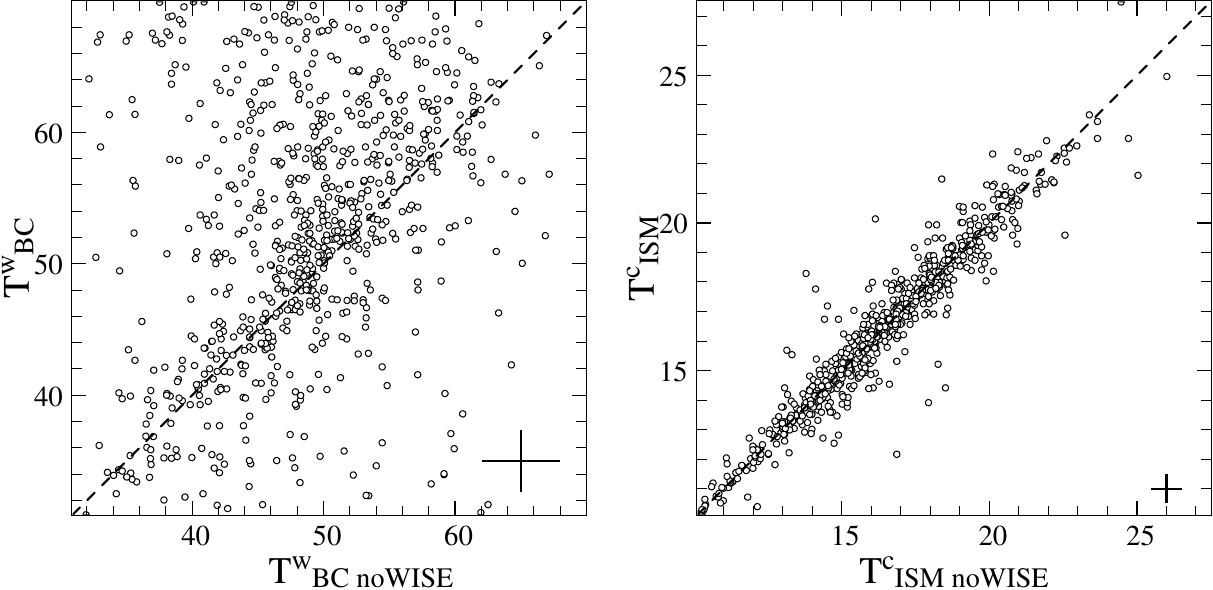}
 \end{center} 
 \caption{Warm dust heated in star forming clouds ($T^W_{BC}$, left panel) and cold dust in thermal equilibrium within the ISM ($T^C_{ISM}$, right panel) parameters obtained with MAGPHYS \citep{dac} using the full photometric coverage (ordinate) or removing WISE data (abscissa). The dashed line shows a linear relation between the two quantities and the black cross in the bottom right corner shows the average errors in the fit.}\label{temp}\end{figure}

These results are independent of the method adopted, as we found similar conclusions using MAGPHYS, which uses a different approach to the problem. \cite{dac} model the IR emission due to the dust as a combination of modified black bodies with different temperatures. With this method, we can estimate the emission of the PAHs, but also the temperatures of the cold dust in the ISM with temperatures in the range 10-30 K, and the warm dust in star-forming clouds, with temperatures of 30-70 K. Figure \ref{temp} shows the temperatures of the cold and warm dust components obtained with the two data sets with MAGPHYS. Large grains in thermal equilibrium emit most of the modified black body radiation produced by cold dust. The exclusion of MIR data from the fit has a negligible effect on this parameter, while the warm dust component, heated in star-forming regions, becomes unreliable. The consistency of the two methods supports the idea that the absence of MIR data in SED fitting processes leads systematically to overestimation of the star formation and dust luminosity by $\sim$ 30\%. An additional cautionary remark is related to the possibility that the results obtained are dependent on the input parameters used for the fit, similarly to what is seen in Fig. \ref{qpah}. However, while the approach used in MAGPHYS for the stellar component is similar to the one used in CIGALE, for the dust component the methodology is quite different, with MAGPHYS applying modified black body spectra, and CIGALE using the physically motivated models of \cite{dra} (see Sect. \ref{method} for further details). The parameters investigated in this section are strongly linked to the dust component; therefore, we consider the agreement between the two tools as an indication of the  reliability of the results. It is reasonable to assume that in broad terms the biases induced by the input parameters are mitigated through the comparison with different modelisations.

\subsection{Removing FIR data}

\begin{figure*}\begin{center}
\includegraphics[clip=,width=.79\textwidth]{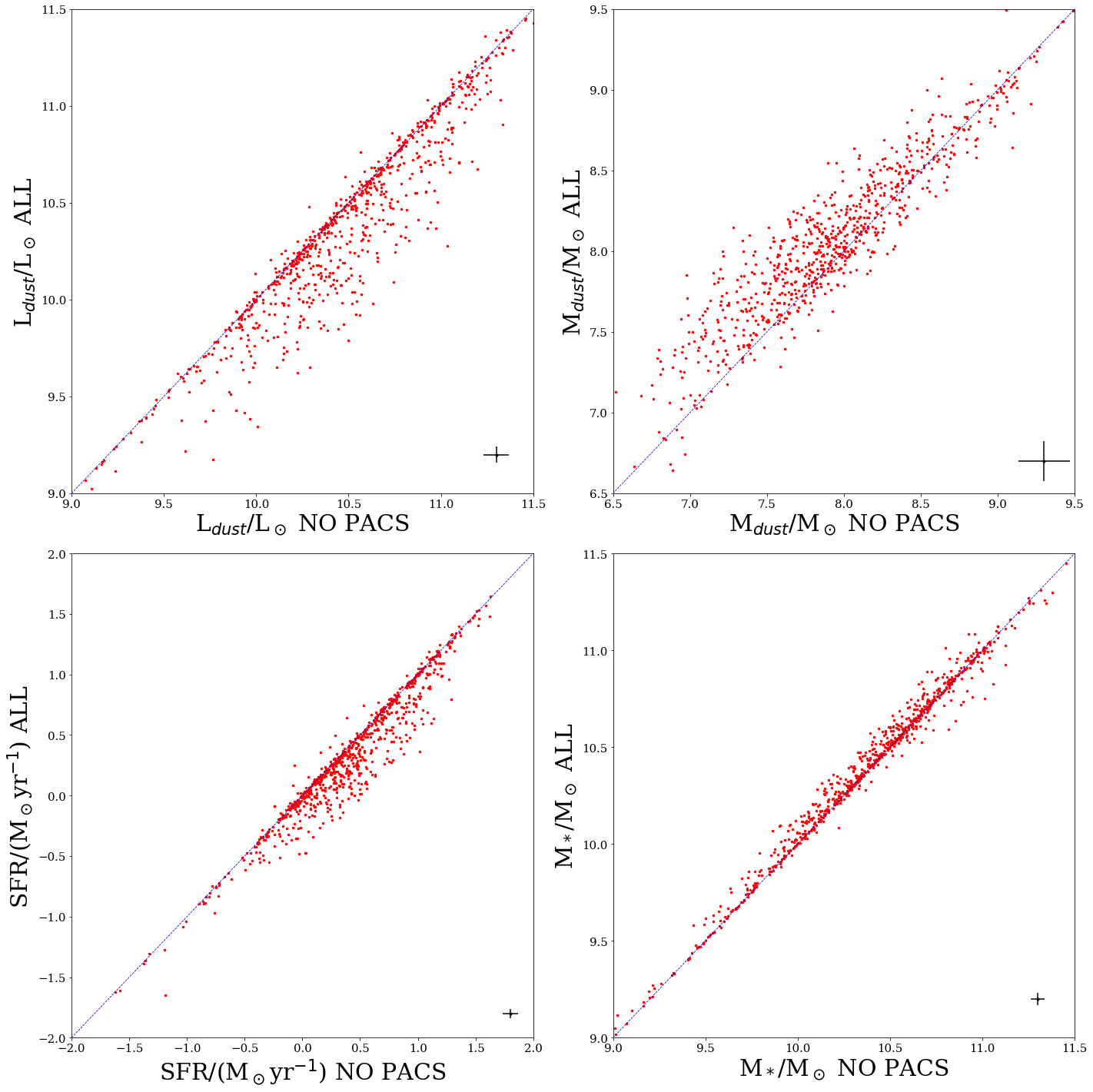}
\includegraphics[clip=,width=.99\textwidth]{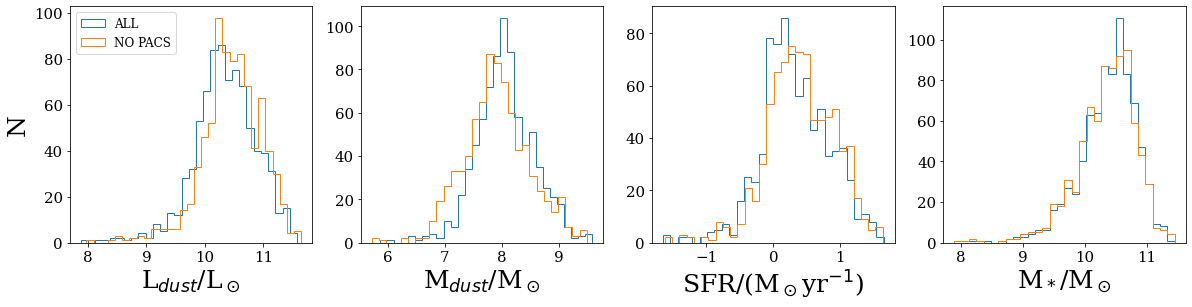}
\end{center} 
\caption{Top panels: Dust luminosity (L$_d$, top left), dust mass (M$_d$, top right), SFR (bottom left), and stellar mass (M$_\ast$, bottom right) obtained with CIGALE \citep{nol,boq} from the sample defined in Sect. \ref{sample} using the full photometric coverage (ordinate) or removing PACS data (abscissa). The blue dashed line shows a linear relation between the two quantities and the black cross in the bottom right corner shows the average errors in the fits. Bottom Panels: Histograms of the same quantities.}\label{nopacs}\end{figure*}

We also investigated the effect of removing PACS data from the full data set, leaving the wavebands between 25 $\mu$m and 200 $\mu$m  unconstrained. The bulk of the dust mass resides in cold dust grains at temperatures in the range of 15-25 K mostly emitting at $>$200 $\mu$m, while the warmer but less massive components radiate at 25-200 $\mu$m \citep{ben2,ben3}. Figure \ref{nopacs} shows that when the Wien side of the cold dust SED is unconstrained, we see negligible variations in stellar mass estimates, while the fits overestimate SFRs by $\sim$30\%, similarly to what is seen for MIR data in Sect. \ref{sec_nowise}. A lack of PACS data leads to overestimation of the dust luminosities by $\sim$25\% and underestimation of the dust masses by $\sim$40\%.

\begin{figure*}\begin{center}
\includegraphics[clip=,width=.3\textwidth]{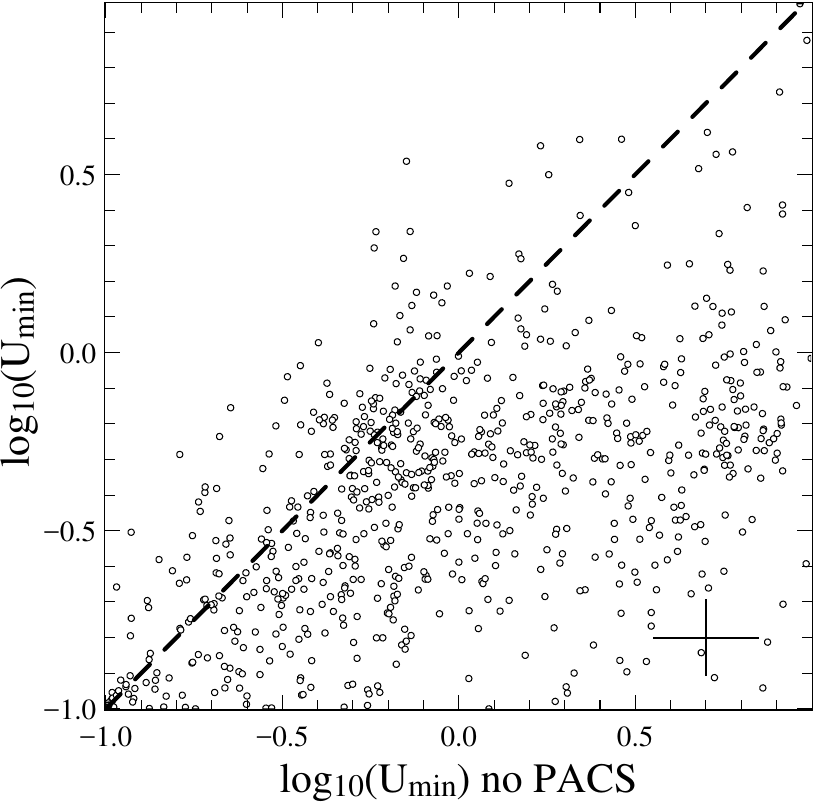}
\includegraphics[clip=,width=.31\textwidth]{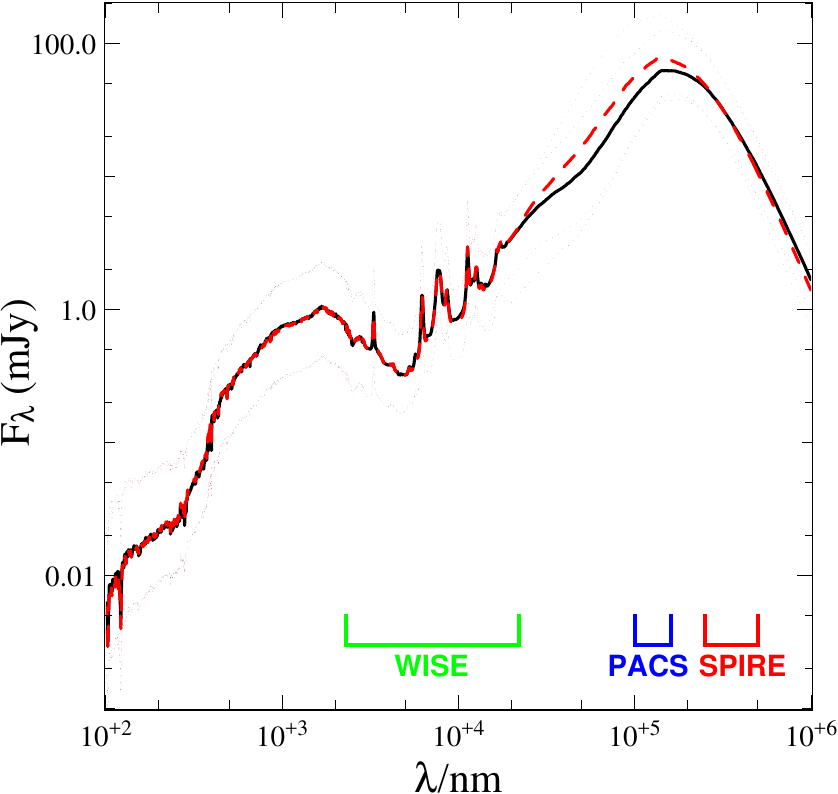}
\includegraphics[clip=,width=.29\textwidth]{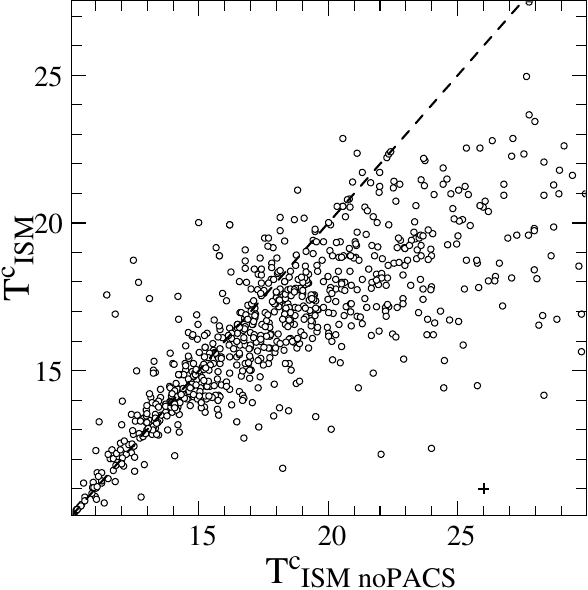}
\end{center} 
\caption{Left panel: $U_{min}$ parameter of \cite{dra} obtained with CIGALE \citep{nol} using the full photometric coverage (ordinate) or removing PACS data (abscissa). The dashed line shows a linear relation between the two quantities and the black cross in the bottom right corner shows the average errors in the fit. Middle panel: Median SED with 16th and 84th percentiles (dotted line) obtained using the full photometric coverage (black solid) or removing PACS data (red dashed line). Right panel: Cold dust in thermal equilibrium within the ISM ($T^C_{ISM}$) parameter obtained with MAGPHYS \citep{dac} using the full photometric coverage (ordinate) or removing PACS data (abscissa). The dashed line shows a linear relation between the two quantities and the black cross in the bottom right corner shows the average errors in the fit.}\label{umin2}\end{figure*}

The lack of constraints in PACS wavebands leads to higher $U_{min}$, as shown in the left panel of Fig. \ref {umin2}. This in turn leads to lower mass estimates, as seen in nearby galaxies \citep[e.g. Fig. 8 in][]{ani}. The middle panel of Fig. \ref {umin2} compares the median SEDs of the two samples and shows how the fits obtained removing PACS data (red dotted line) are systematically higher than the SEDs from the full photometric analysis. The peak is also slightly shifted towards lower wavelengths, producing higher dust temperatures. 

The MAGPHYS analysis confirms this trend, as shown in the right panel of Fig. \ref{umin2}. The cold component of the dust emission, with temperatures above 15 K, is systematically overestimated when removing PACS data. As most of the dust mass is at temperatures in the range of 15-25 K, the fits overestimate the dust mass in thermal equilibrium, as seen in the top right panel of Fig. \ref{nopacs}. The lack of PACS data has a twofold effect: it affects the dust luminosities through inaccurate starlight intensity estimates and alters the dust masses because of misleading estimates of the interstellar radiation field, as shown in Fig. \ref{nopacs}.

\section{The interplay between stellar and dust emission}
\label{discussion}

The methods described in Sect. \ref{method} rely on the assumption that the amount of radiation produced in star-forming and nebular clouds is partially absorbed by dust and then re-emitted at IR wavelengths.  The consequence of this absorption process is an expected correlation between the dust luminosity and the SFR. Observations confirm this trend, leading to the definition of different SFR tracers based on IR emission \citep[see][for a complete review]{ken1,ken2}. These relations assume completely dust-obscured  star formation and dust heating processes dominated by young stars, but these assumptions require more thorough examinations in light of recent results \citep[e.g.][]{gal1,gal2}. In the following sections, we analyse these correlations in our sample, showing how a multiwavelength approach is necessary in order to overcome the systematic effects induced by the above-mentioned assumption.

\subsection{Ultraviolet absorption in the SFR--dust luminosity correlation}

Infrared-based SFR calibrators have been calibrated in nearby spiral galaxies, assuming that the dust obscures a high fraction of the emission produced during star formation. As suggested by observations ---with various instruments--- of UV emission from star-forming regions that is not absorbed by dust, this assumption must be taken with caution \citep{hir}. The dust luminosity correlates more tightly with the SFR in galaxies that are not detected at UV wavelengths, as in strong starburst galaxies, for example. For galaxies with moderate SFRs of between 1 and 10 M$_\odot$ yr$^{-1}$, the correlation between L$_{d}$ and SFR is more scattered \citep{smi}, implying a more variable fraction of absorbed UV radiation. This physical mechanism is quantified by FUV attenuation, $A(FUV)$, which parametrises the fraction of UV radiation emerging from star-forming clouds absorbed by dust.

The correlation between the dust luminosity and the SFR is tighter for dustier galaxies, where the dust obscuration dominates \citep[see also][]{cle,pap2}; although at lower luminosities the contribution from old stars to dust heating can be relevant, impacting its reliability as an SFR tracer. \citealt{cle} investigated the dust properties of 234 local star-forming galaxies sampling their SEDs in the range 0.36 $\mu$m $< \lambda <$ 1381 $\mu$m. These authors found that for high L$_{d}$, the total IR luminosity is a good proxy for the SFR except for regions where L$_{d} <$ 5 $\times$ 10$^9$ L$_\odot$. They argue that, in certain objects, the IR emission of the warm dust would provide a good estimate of the SFR, but that in other, less extinct objects the optical and UV emission would also need to be taken into account \citep[see Fig. 9 in][]{cle}.

\begin{figure}\begin{center}
\includegraphics[clip=,width=.4\textwidth]{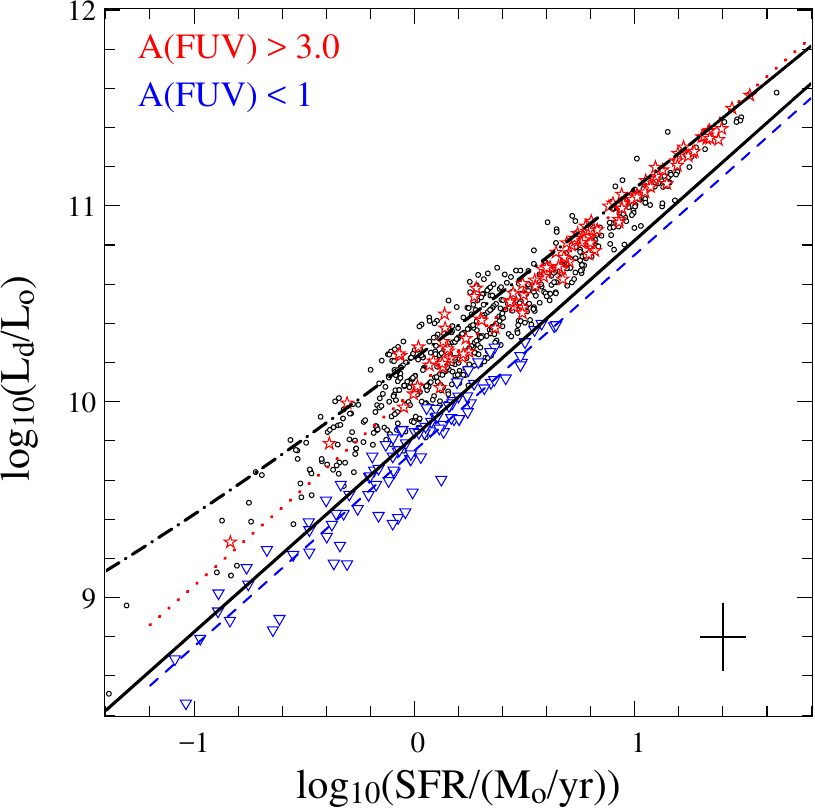}
\includegraphics[clip=,width=.4\textwidth]{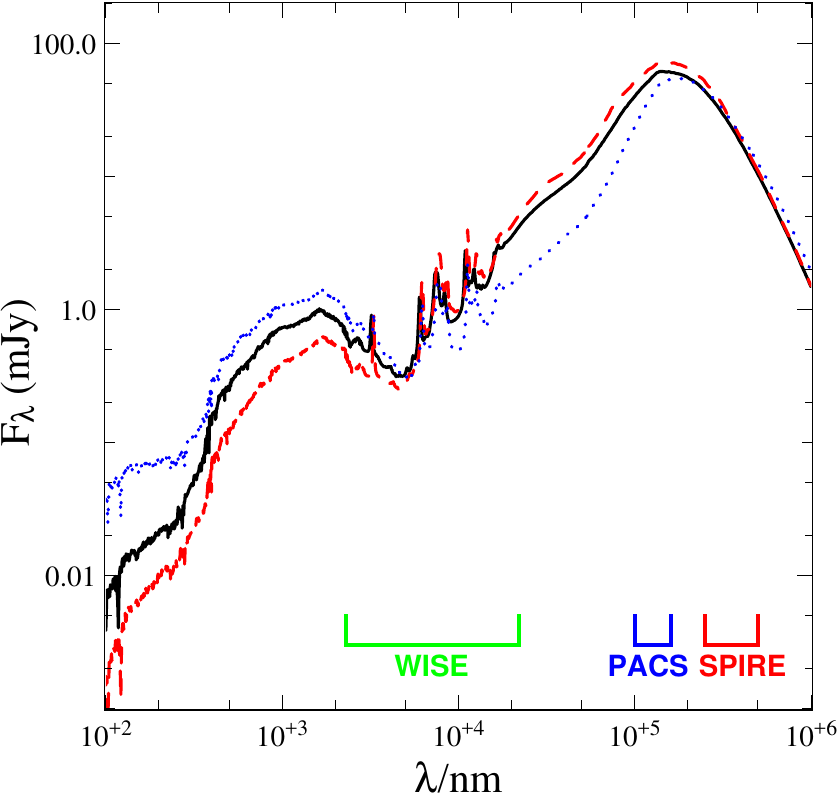}
\end{center} 
\caption{Top panel: Dust luminosity ($L_d$) as a function of the SFR obtained with CIGALE \citep{nol} using the full photometric coverage. Red stars and blue triangles identify galaxies with $A(FUV) >$ 3.0 and $A(FUV) <$ 1, with their best fit shown as red dotted and blue dashed lines, respectively. Red dotted and blue dashed lines show the best fit for the galaxies showing high and low levels of extinction, respectively. The solid and dot-dashed black lines show the SFR as derived from Eqs. \ref{eqken} \citep{ken1} and \ref{eqcle} \citep{cle}, respectively. Bottom panel: Median SED obtained using the full photometric coverage (black solid line), considering the galaxies with $A(FUV) >$ 3.0 (red dashed line) and $A(FUV) <$ 1 (blue dotted line). The photometric coverage of WISE, PACS, and SPIRE bands is shown at the bottom.}\label{ldsfr}\end{figure}

If the scatter at low dust luminosities is due to a low galaxy extinction, separating galaxies according to the FUV attenuation would end up selecting galaxies where the dust luminosity correlates with SFR. To investigate this hypothesis, we divided our sample into galaxies with low and high extinction, $A(FUV) <$ 1) and $A(FUV) >$ 3, respectively. These values correspond to the median dust attenuation in FUV measured for galaxies selected in the NUV or FIR \citep{bua,bur}, allowing a rough estimation of the bias induced by the different selection criteria. The top panel of Fig. \ref{ldsfr} confirms that by adding $A(FUV)$ as a third dimension in the $L_d-$SFR plane, we select families of galaxies for which the scatter in the correlation is significantly reduced. In Fig. \ref{ldsfr} we also show the best fit for the two subsamples (dashed and dotted lines), and we compare them to the SFR calibrators defined in \citealt{ken1}:

\begin{equation}
\textrm{SFR [M$_\odot$ yr$^{-1}$] = 1.5 $\times$ 10$^{-10}$ L$_{IR}$ [L$_\odot$]},
\label{eqken}
\end{equation}

and \citealt{cle}:

\begin{equation}
 \textrm{SFR} \ [\textrm{M$_\odot$ yr$^{-1}$}] = 10^{\Big[-9.6+\textrm{log}\frac{L_d}{L_\odot}-\Big(\frac{2.0}{\textrm{log}\frac{L_d}{L_\odot}-7.0}\Big)\Big]}.\label{eqcle}
\end{equation}

The SFR calibrator defined in \cite{ken1} fits the star-forming galaxies with low dust extinction  better, while the calibrators of \cite{cle} fit  the galaxy with high $A(FUV)$ and L$_d >$ 10$^{10}$ L$_\odot$ very well. These differences are related to the sample adopted to define the correlation: both samples consider galaxies at distances below 150 Mpc, but while \cite{ken1} recover Eq. \ref{eqken} from the work of \cite{hao}, selecting galaxies from the SINGS sample \citep{ken3}, \cite{cle} select their objects from the Planck Early Release Compact Source Catalogue \cite[ERCSC,][]{pla}. Equation \ref{eqken} is calibrated for local star-forming galaxies with SFR between 0 and 15 M$_\odot$ yr$^{-1}$ and L$_d$ between 10$^7$ and 10$^{11}$ L$_\odot$, considering relatively dust-poor objects. Planck selection privileges dust-rich galaxies, where even for low SFR values the UV attenuation is high. This once more  underlines the need for a multi-wavelength approach to quantify these shifts in the SFR-L$_d$ calibration due to UV absorption: we find that attenuation of 2 mag in the FUV produces an increase of $\sim$50\%  in the dust luminosity independently of the shape of the star formation history recovered through the fit. The increased attenuation renormalises the SFR--L$_d$ correlation towards higher L$_d$ (red dotted and blue dashed line in Fig. \ref{ldsfr}), a trend confirmed through the comparison with the works of \cite{ken2} and \cite{cle}. For each of the subsamples defined in Fig. \ref{ldsfr}, the amount of UV radiation absorbed is comparable, despite the different values of dust luminosities or SFR. 

The dust-heating mechanism for low-redshift galaxies in the main sequence can be highly composite, involving both the star-formation process and the radiation field produced by stars in late evolutionary stages \citep{lon,ben,ben2,ben3,via,via2,boq2,ner}. This combined effect alters the reliability of the dust luminosity as a star formation tracer, at least at lower SFRs \citep[see e.g.][]{ken2,pap2}. Adding a further constraint on $A(FUV)$ we tighten the correlation, selecting families of galaxies for which the dust-heating mechanisms are mostly similar. The correlation between SFR and dust luminosity is still valid, but as a consequence of the variations in the dust-heating mechanism the y-intercept shifts to higher or lower values according to the fraction of UV radiation that the dust can absorb, as shown by dashed and dotted lines in Fig. \ref{ldsfr}.

Increasingly attenuated galaxies  have on average higher dust luminosity and higher star formation rate (Fig. \ref{ldsfr}). The median SED, shown in the right panel of Fig. \ref{ldsfr}, confirms this trend and shows that higher FIR fluxes are found for highly attenuated galaxies. The MIR fluxes in the central regions of SED fits are quite different, underlining the importance of those wavelengths in SED fitting methods \citep{bua}. 

\subsection{The link between dust mass and SFR}

\begin{figure}\begin{center}
\includegraphics[clip=,width=.35\textwidth]{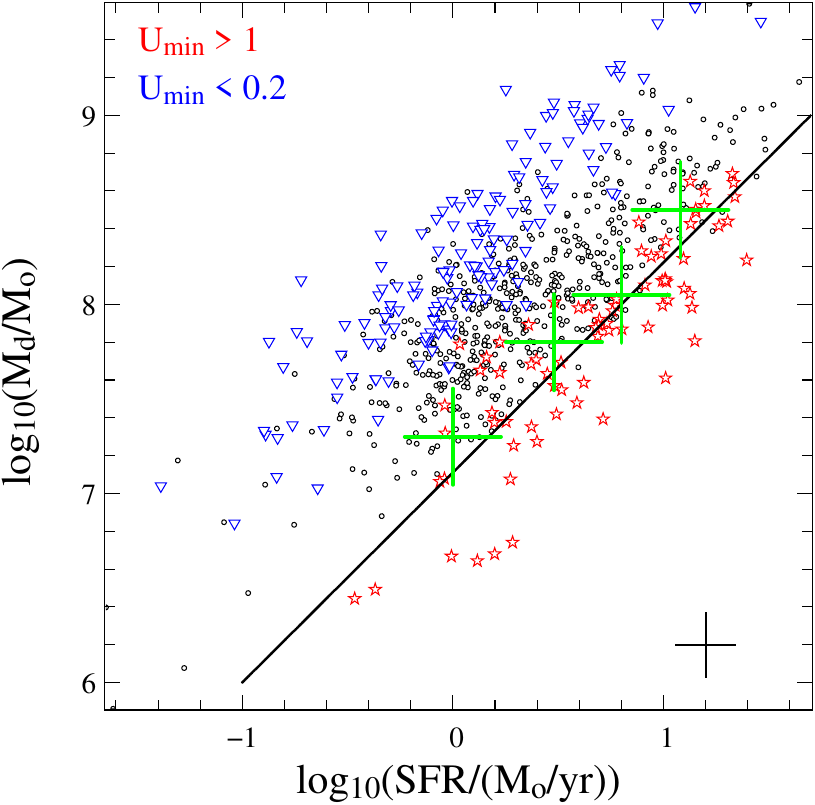}
\includegraphics[clip=,width=.35\textwidth]{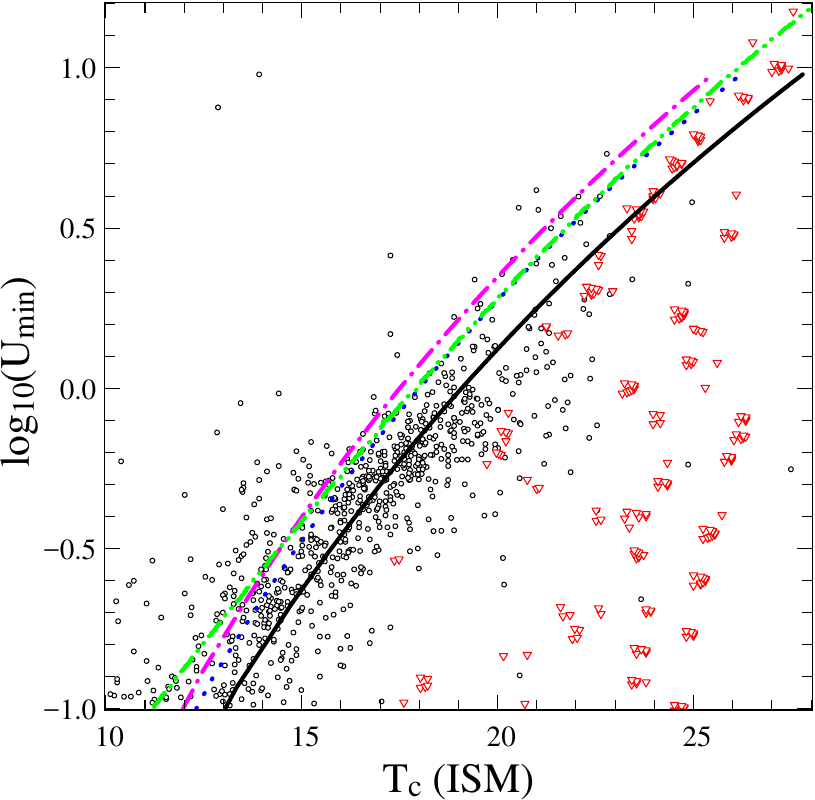}
\end{center} 
\caption{Top panel: Dust mass ($M_d$) as a function of the SFR obtained with CIGALE \citep{nol} using the full photometric coverage. Red stars and blue triangles identify galaxies with $U_{min} >$ 1.0 and $U_{min} <$ 0.2, respectively, and the black cross in the bottom right corner shows the average errors in the fit. The solid line and the green crosses show the results from \cite{dac2} and \cite{smi}, respectively. Bottom panel: Black circles show the $U_{min}$ obtained with CIGALE as a function of the temperature of the cold dust in thermal equilibrium within the ISM ($T^C_{ISM}$) obtained with MAGPHYS using the full photometric coverage. Red triangles show the results obtained with \cite{dra} models in \cite{hun} for KINGFISH. Magenta dot-dashed, green double dot-dashed, and blue dotted lines show the best fit obtained by \cite{boul} for the solar neighborhood with T$_d$ = A $\cdot U_{min}^{1/(4+\beta)}$ (A = 17.5 K and $\beta$ = 2), the best fit obtained with \cite{dra} models in KINGFISH \citep{hun}, and the relation of \cite{dra} with A = 18 K and $\beta$ = 2 (see also Eq. 13 in \citealt{dra3}), respectively. The black solid line shows the best fit for our sample obtained for A = 19.1 K and $\beta$ = 2.}\label{md_sfr}\end{figure}

Investigation of the dust properties in a galaxy requires the conversion of the dust luminosity into dust mass, which is a relatively  complex problem. As mentioned earlier, the use of IR emission as an SFR tracer finds its justification in the assumption of a dust heating process dominated by young stars. Several studies have shown that at low SFRs a non-negligible fraction of dust can be heated by the radiation field produced by stars in late evolutionary stages \citep{lon,ben,ben2,ben3, via2,via,boq2}. Furthermore, the correlation between the SFR and dust mass is weaker than that  between
the SFR and dust luminosity \citep{cle,smi,pap2} because the bulk of dust mass resides in cold grains with temperatures of around 15-25 K. The dust mass is almost certainly related to SFR through the Kennicutt-Schmidt relation. However, the dust mass likely traces a combination of atomic and molecular gas \citep{pap3}, while SFR depends primarily on the molecular gas mass \citep[e.g.][]{big}, causing the scatter seen in Fig. \ref{md_sfr}. The warm component, although a small fraction, has a higher luminosity in star-forming galaxies, affecting the relation in Fig. \ref{md_sfr}. The dust heated by $U_{min}$ is quantified in \cite{dra} models through $\gamma$ (dust mass fraction in star-forming regions), with average values of around a few percent \citep{bia}, but the dust in thermal equilibrium has a different emissivity according to its temperature, and the two quantities suffer a high level of degeneracy \citep{juv}. However, the variation of $\gamma$  mainly affects the SED for 10 $< \lambda <$ 60 $\mu$m \citep[Fig. 18 of][]{dra}, while changes in $U_{min}$ affect a larger region of the SED, between 20 and 1000 $\mu$m \citep[Fig. 13 of][]{dra}. For this reason, our homogeneous coverage at those bands using WISE, PACS, and SPIRE data allows better characterisation of the dust emission, partially breaking this degeneracy and supporting the reliability of the results shown.

Finally, the SFR is an extensive quantity (i.e. it depends on the size of the system), while $U_{min}$ is an intensive quantity that depends on the geometry. Two galaxies with the same SFH and dust mass but with different dust geometries will have different $U_{min}$,  depending, for example, on the galaxy compactness. Nevertheless, the dust heated by the interstellar radiation field remains the dominant component when estimating dust masses, mostly because it fixes the cold dust temperature. It is not perfect but good enough for a first cut, and for this reason, the correlation between dust mass and the SFR is tighter if we select galaxies according to their $U_{min}$ values. The top panel of Fig. \ref{md_sfr} shows that galaxies with a given SFR have larger dust masses if $U_{min}$ is lower. If $U_{min}$ is lower, a higher mass of dust is needed to produce a certain dust luminosity given the SFR and the fraction of photons absorbed (and therefore re-emitted because of the energy balance). As a consequence, the subsamples defined according to $U_{min}$ show a stronger correlation than the overall sample.

Figure \ref{md_sfr} shows the best fit obtained with IRAS-selected SINGS galaxies \citep{dac2}, and the 250 $\mu$m-selected objects in H-ATLAS \citep{smi}. Both surveys seem to be more consistently fitted with values of $U_{min} >$ 1 (red stars in the top panel of Fig. \ref{md_sfr}). This result is consistent with those of \cite{pap2}, who show that IRAS-selected galaxies tend to be hotter than those selected by {\it Herschel} \citep[see also][and their analysis]{smi}, and also that with the eight cross scans of the SPIRE instrument, HeViCS detect dusty galaxies with lower FIR emission, investigating the region covered by galaxies with U$_{min} <$ 0.2 (blue triangles in Fig. \ref{md_sfr}).

\subsubsection*{Interstellar radiation field and cold dust temperature}

Finally, we discuss the correlation between the interstellar radiation field and the temperature of dust in thermal equilibrium, shown in the bottom panel of Fig. \ref{md_sfr}. The correlation between these two quantities was also discussed by \cite{ani}, who showed that when modelling the \cite{dra} SEDs with a blackbody multiplied by a power-law opacity, the dust temperature is proportional to $U_{min}$. In Fig. \ref{md_sfr} we also show the best fit obtained in \cite{boul} for the solar neighbourhood with the DIRBE and FIRAS experiment given by T$_d$ = A $\cdot U_{min}^{1/(4+\beta)}$ (A = 17.5K, and $\beta$ = 2), the fits to the KINGFISH data applying \cite{dra} models, and the relation adopted by \citet{dra} with A = 18 K and $\beta$ = 2 \citep[see Eq. 13 in][]{dra3}. Our sample shows an emissivity consistent with a value of $\beta$ = 2, but the best fit recovers an average dust temperature of 19.1 K, slightly higher than what was measured in the solar neighbourhood (black line in Fig. \ref{md_sfr}). This is caused by the dust temperatures, which are warmer at higher redshift \citep{dun}, as shown by the fit obtained by \cite{hun} for nearby galaxies of KINGFISH (red triangles in Fig. \ref{md_sfr}).

\section{The promise of JWST}
\label{jwst}

Until now, observations at MIR wavebands have not been sufficient to explore Universe at $z >$ 2 with sufficient statistics. Surveys in well-known fields, such as the Cosmic Assembly Deep Near-Infrared Extragalactic Legacy Survey \citep[CANDELS,][]{ash,gro2,koe} and the Great Observatories Origins Deep Survey \citep[GOODS,][]{gia,wan,hat2,lin} performed with the Spitzer Space Telescope \citep{wer}, have enormously increased our knowledge of the high-redshift Universe, but substantial uncertainties remain. However, these wavebands are fundamental in constraining the stellar radiation field and the obscured star formation. A leap forward in this area will occur in the future, with the launch in 2021 of the James Webb Space Telescope \citep[JWST,][]{gar}. The nine broad-band filters one of its onboard instruments, the Mid-Infrared Instrument \citep[MIRI,][]{rie2,wri2}, will cover the wavelength range 5-28 $\mu$m contiguously, with a PSF FWHM of $\sim0.2''-0.8''$, and a sensitivity of one to two orders of magnitude higher than Spitzer \citep{gla}. With a field of view of 1.3$' \times 1.7'$, MIRI has the potential to revolutionise the field, investigating galaxies during the reionisation epoch at $z >$ 6 (NIR rest frame).

In this paper and in \cite{pap2}, we identified a sample of galaxies with moderate SFR $\sim$ 2 M$_\odot$yr$^{-1}$ populating the main sequence at $z < 0.5$. This sample represents dusty star-forming galaxies with stellar masses of M$_\ast \sim$ 3$\times10^{10}$ M$_\odot$, which at a median redshift of $z = 0.1$ populate the knee of the stellar mass density \citep{fur}. This range of masses is also important for characterising the mass assembly history of galaxies because, as shown by \cite{ilb2}, galaxies with M$_\ast \sim$ 1-3 $\times 10^{10}$ M$_\odot$ have experienced continuous growth in specific star formation rate (sSFR) between $z = 1$ and $z = 4$. In this respect, it is interesting and instructive to test if and how the MIRI instrument will detect the SEDs obtained from our fits. This extrapolation from low-$z$ galaxies, or using semi-analytical models, is fundamental to quantify  the number of sources detected on
one hand, and their physical properties on the other \citep[see also][]{mas,furl,cow,wil,yun}. 

Reasonable doubts could arise regarding the reliability of these tests. After all, we are considering a sample of low-redshift SEDs as representative of galaxies in the early Universe, up to $z$ = 8, less than 1 Gyr after the big bang. The conditions of the Universe at those ages were different, and galaxies were more complex, with substructures, clumpy emission, and variable sizes. However, we consider our exercise meaningful mainly for two reasons: firstly, the evolution of the galaxy sizes with redshift is not clear and depends significantly on the definition of the isophotal limits. Assuming symmetric structures, \cite{shi} for example found that size decreases with redshift, but using more adapted isophotal limits, which take into account the structure complexity mentioned above, \cite{rib} found constant sizes in the range 2 $< z <$ 4.5. On the other hand, because of its clumpy
structure a galaxy  could be detectable in a specific filter thanks to the increased emission at those wavelengths \citep[e.g.][]{sob}. In this sense, lower surface brightness associated with extended multiple components in galaxies does not necessarily exclude detection in one of the MIRI filters.

\begin{table}\centering
Sensitivity of MIRI filters
\begin{tabular}{c|c|c}
\hline
Filter &  Sensitivity at S/N=10 & CEERS\\
       &   $\mu$Jy & $\mu$Jy\\
          (1)  &  (2) & (3)\\
        \hline
        F560 & 0.13    & \\
        F770 & 0.24    & 0.23\\
        F1000W & 0.52  & 0.43\\
        F1130W & 1.22  & \\
        F1280W & 0.92  & 0.7\\
        F1500W & 1.45  & 1.1\\
        F1800W & 2.97  & 2.5\\
        F2100W & 5.14  & 2.75\\
        F2550W & 17.3  &\\
        \hline
    \end{tabular}
    \caption{Sensitivity of MIRI filters in 10 ks at S/N = 10 and the Cosmic Evolution Early Release Science Survey \citep[CEERS,][]{fin}. 1) MIRI filters; 2) Detection limit at S/N = 10 in 10ksec; 3) 5$\sigma$ sensitivity for CEERS program.}\label{miri}
\end{table}

\begin{table}\centering
Detection rate (\%)
\begin{tabular}{c|c|c|c|c}
\hline
MIRI Filter &  z = 2 & z = 4 & z = 6 & z = 8 \\
\hline
F560   & 99 & 94.6 & 82.9  & 49.7\\
F770   & 96 & 90.7 & 69.5 & 44.5\\
F1000   & 86.5 & 70 & 44.3 & 22.1\\
F1130   & 52.7 & 30.4 & 15.4 & 2.6\\
F1280   & 61.7 & 27.4  & 23.3 & 8.2\\
F1500   & 34.6 & 9.2 & 7.3 & 3.6\\
F1800   & 27.8 & 0.9 & 0 & 0\\
F2100   & 25.9 & 0 & 0 & 0\\
F2550   & 13.5 & 0 & 0 & 0\\
\hline
\end{tabular}
\caption{Detection rate with MIRI filters at different redshifts.}\label{detectionrate}
\end{table}

\begin{figure*}\centering
    \includegraphics[clip=,width=.99\textwidth]{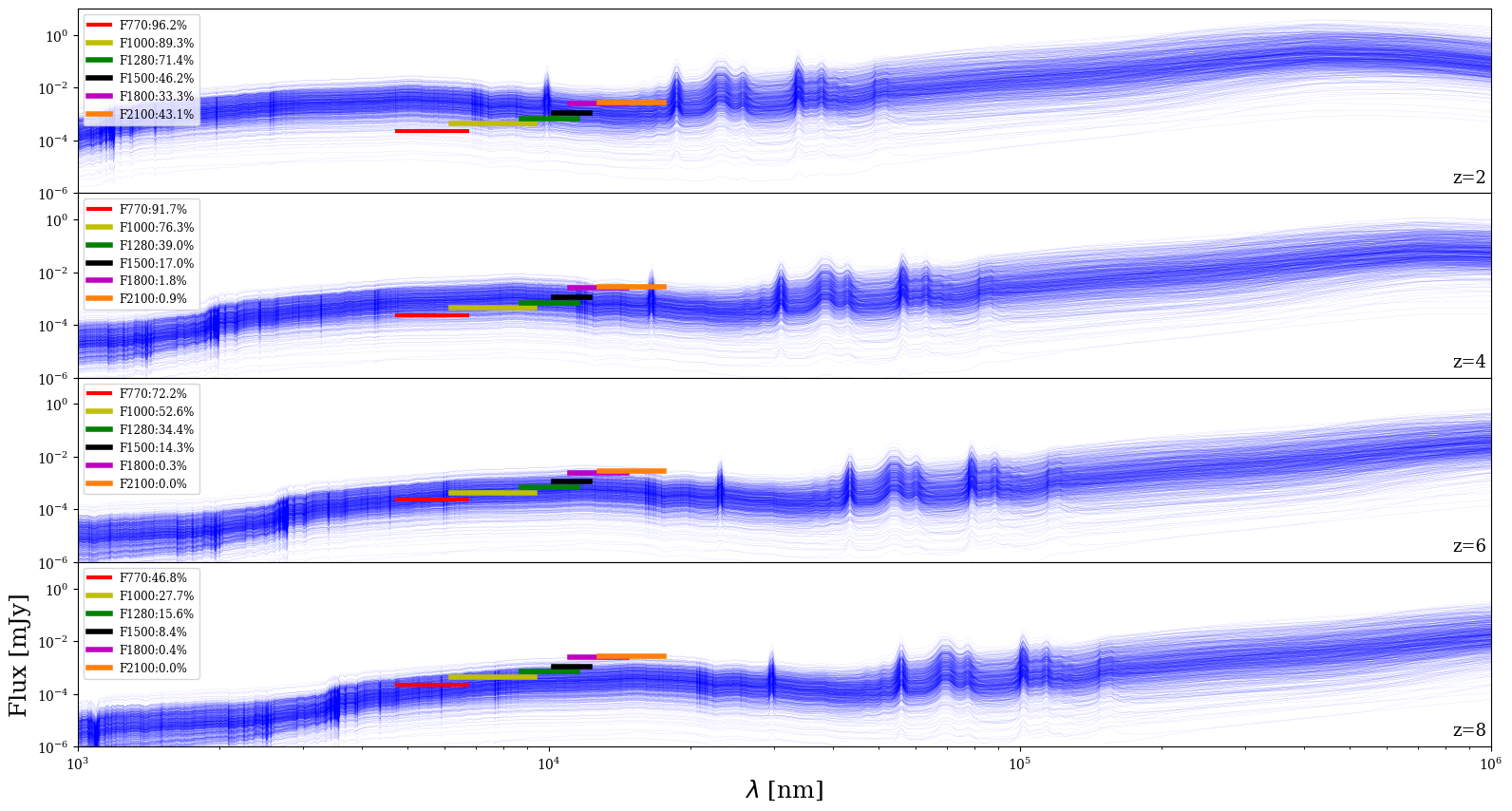}
    \caption{CIGALE best-fit SEDs of our sample at different redshifts, marked in the bottom right of each panel. Horizontal coloured bars show the wavelength coverage and the sensitivity of MIRI filters.}\label{miriTotalSample}
\end{figure*}

As a first test, we considered the minimum detectable flux densities for a signal-to-noise ratio (S/N) = 10 in 10ksec for an unresolved point source, as reported in Table \ref{miri} based on \cite{gla} and \cite{pon}. The detection rate of the sample at redshifts of 2, 4, 6, and 8 is reported in Table \ref{detectionrate} and shown in Fig. \ref{miriTotalSample} together with the MIRI filter coverage. At $z$ = 2, more than 60\% of the galaxies in our sample are detected with at least five filters, sampling the region of PAH emission. At higher redshift, the filters below 15 $\mu$m (F560, F770, F1000) guarantee a detection rate of around 50\% up to $z$ = 6. As expected, at $z$ = 8 the detection rate falls drastically, however for $\sim$45\% of the sample we still have detection in two filters, F560 and F770, and for 22\% we  also have a detection with F1000. A promising result from this statistical study is that for a subset of $\sim$8\% of the sample, we will also have a detection with F1280, together with F560, F770, and F1000, covering  the range 4-15 $\mu$m (observed wavelengths) homogeneously. This photometric coverage will improve the accuracies of the fits \citep{bis1,bis2,kau}, allowing a more precise continuum estimation in SED regions dominated by the emission of the evolved stars (t > 10$^8$ yr), which form the bulk of the stellar mass. This in turn will put stronger constraints on the stellar mass function at those redshifts \citep[see e.g.][]{fur,ilb2}, shedding light on the mechanism of mass assembly in the early Universe.

This test refers to sensitivity values estimated in the best possible conditions, which are useful for quantifying the instrument capabilities. In the following, we repeat our test considering more realistic values obtained from an approved JWST program, the Cosmic Evolution Early Release Science Survey \citep[CEERS,][]{fin}, which will observe 100 arcmin$^2$ in the Extended Groth Strip Field (EGS). The aim of this project, which is one of the first publicly available JWST surveys, is to observe this region with four pointings and different filters. One of these pointings, MIRI2, will observe EGS with F770, F1000, F1280, F1500, and F1800, for 1665 seconds/band, and F2100 for 4662 seconds reaching a 5$\sigma$ sensitivity of 0.23 (F770), 0.43 (F1000), 0.7 (F1280), 1.1 (F1500), 2.5 (F1800), and 2.75 (F2100) $\mu$Jy, reported in Table \ref{miri}.

\begin{table}    \centering
    \begin{tabular}{c|c|c}
    \hline
        Samples &  SFR & M$_\ast$\\
                &  M$_\odot$yr$^{-1}$ & 10$^{10}$M$_\odot$ \\
            (1) &  (2) & (3)\\ 
            \hline
         Total  & 1.85 & 2.7 \\
         Low    & 0.98 & 1.3 \\
         High   & 4.46 & 4.9 \\
         \hline
    \end{tabular}\caption{Median values of star formation rate (col. 2) and stellar mass (col. 3), for the samples reported in col. 1: 'Total' refers to the total sample, while 'Low' and 'High' refers to subsamples with values below or above the median, respectively. Units are reported in the second row.}\label{median}
\end{table}

\begin{figure*}    \centering
    \includegraphics[clip=,width=0.9\textwidth]{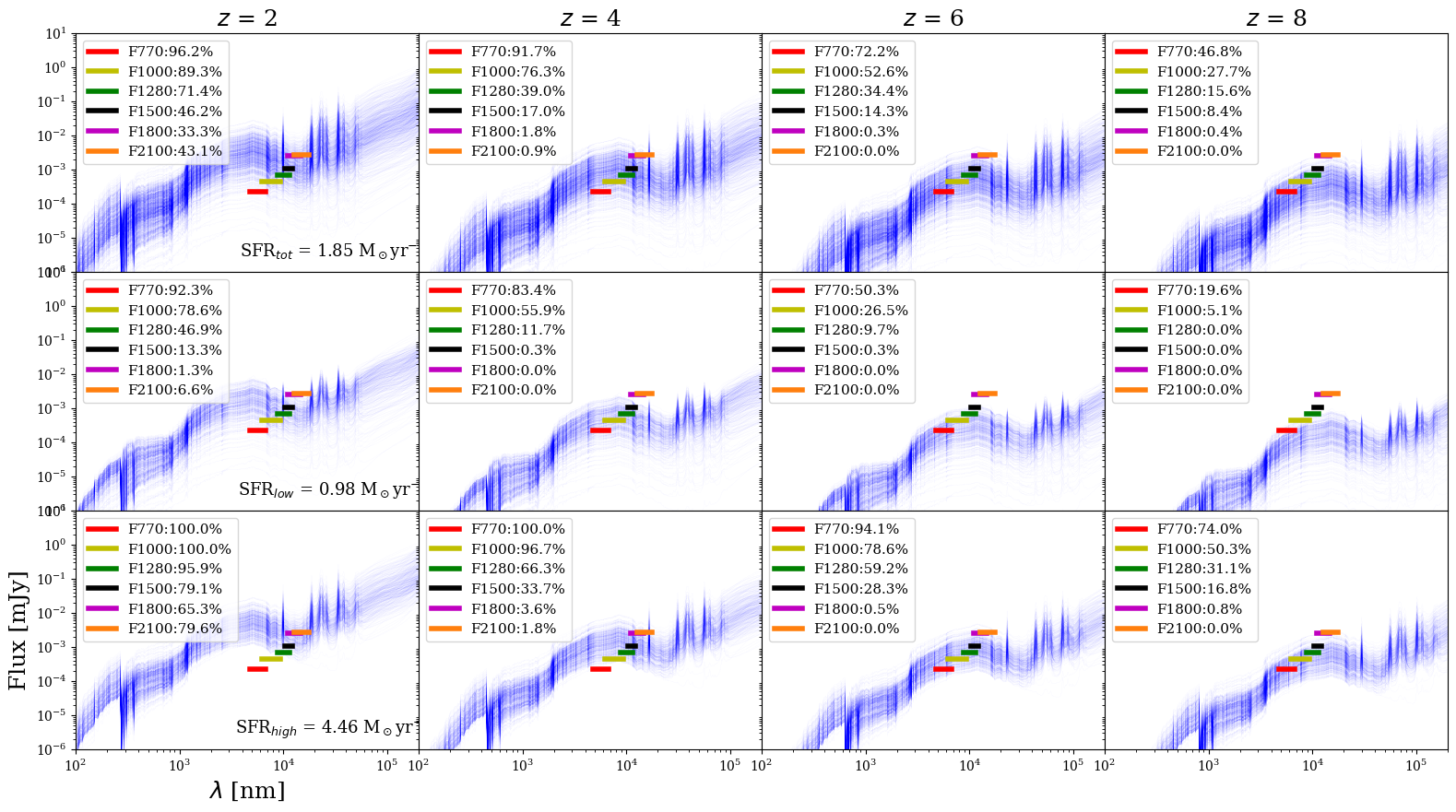}
    \includegraphics[clip=,width=0.9\textwidth]{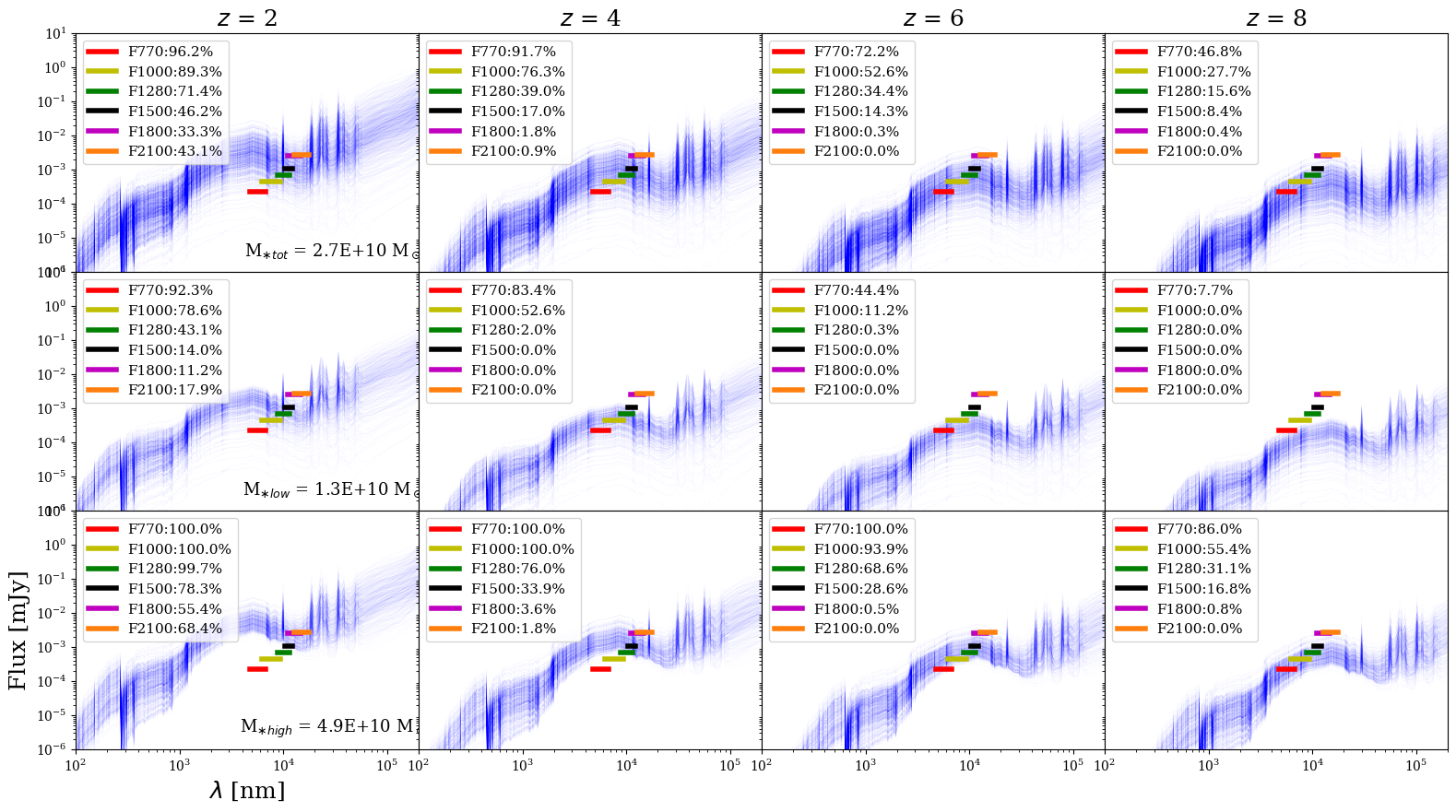}
    \caption{Top panels: Best-fit SEDs of CIGALE at different redshifts, marked on top of each column. Coloured horizontal bars indicate the MIRI filters planned for the CEERS survey. Boxes on the left of each panel show the percentage of detection at a given filter. Galaxies are classified according to the SFR with the starting sample on the top row, the `low' subsample in the middle, and the `high' subsample on the bottom row. Median values of each sample are reported in the bottom right of the first panel in each row and Table \ref{median}.
    Bottom panel: Same analysis selecting galaxies according to their stellar masses, as reported in Table \ref{median}.}\label{sfrCEERS}
\end{figure*}

We considered SFR and stellar mass (M$_\ast$) for four redshift bins: 2, 4, 6, and 8. We calculated the median for each parameter, defining two subsets of galaxies above or below those values. Table \ref{median} reports the median of the total sample (tot) and the two subsamples (low, high). The objects below the median are of great interest in this exercise: galaxies in the knee of the stellar mass function, with SFR $\sim$ 1 M$_\odot$yr$^{-1}$ and M$_\ast \sim 1\times10^{10}$ M$_\odot$ are in the `low' subsample of Table \ref{median}, similarly to dust-poor objects, typically less luminous than local LIRGs (L$_d \sim 10^{11}$ L$_\odot$). The top panel of Fig. \ref{sfrCEERS} shows the detection percentage as a function of SFR for the total, low, and high subsamples, as a function of redshift (top panel of each column).

Even for the F770 filter, with a 5$\sigma$ level of 0.23 $\mu$Jy, the detection rate decreases by more than half from $z = 2$ to $z = 8$. The same rate is not followed by the other filters, because the stellar continuum rises more steeply at higher redshift, highly reducing the percentage. For this reason, for the filter immediately close to F770, which is F1000 (5$\sigma$ = 0.43 $\mu$Jy), the percentage goes from 89\% at $z$ = 2 to 28\% at $z$ = 8. As expected, this effect is more relevant for galaxies with low SFR, where, for $z \ge$ 4, filters above F1500 will have almost no detection. This implies that for galaxies with SFR $\sim$ 1 M$_\odot$yr$^{-1}$ already at $z$ = 6 we will see 50\% detection with F770, decreasing to 20\% at $z$ = 8. For such galaxies, only 5\% of the subsample will be detected with F770 and F1000 at $z$ = 8, and only 10\%  with F770, F1000, and F1280 at $z$ = 6. For galaxies with higher SFR, detection with these three filters will be possible up to $z$ = 6 in $\sim$60\% of the subsample, which is quite promising in terms of statistics.

The bottom panel of Fig. \ref{sfrCEERS} shows the same analysis, selecting galaxies according to their stellar mass, as reported in Table \ref{median}. The first result that catches our attention is that for objects with 4.9$\times10^{10}$ $< M_\ast <$ 1.3$\times10^{10}$ the galaxy detection decreases steeply at $z \ge$ 4, with F770 and F1000 detections for $\sim$ 52\% of the sample at $z$ = 4, and 11\% at $z$ = 6. At the highest $z,$ we recover less than 10\% of the subsample with only F770. For the high-mass subsample on the other hand, the detection rate is very high up to $z$ = 6 where we have detection in at least three filters (F770, F1000, F1280) for $\sim$ 68\% of the subsample.

\section{Conclusions}
\label{conclusions}

This paper investigates fundamental correlations that link the dust component with the star formation in a galaxy. The first problem we address is the impact of a lack of MIR and/or FIR data in SED fitting methods. We selected a sample of FIR galaxies lying in the main sequence at $z <$ 0.5 with full photometric coverage between UV and FIR in order to quantify the differences due to the lack of a specific waveband in the fitting process. The main results are as follows:

\begin{itemize}
\item[-] Best fits of stellar and dust mass estimations are relatively unaffected by the lack of data in the WISE range, while SFR and dust luminosity are overestimated by $\sim$20\%-30\%. These differences are not due to misleading PAH emissions, but are related to unconstrained estimations of the stellar radiation field.
\item[-] When removing PACS data, the recovered stellar masses are consistent with the full data set, overestimating the SFR by $\sim$30\%. An interesting result is the overestimation of dust luminosities by $\sim$25\% and the underestimation of dust masses by $\sim$40\%. 

\item[-] In our sample, the dust luminosities and the SFRs show a stronger correlation for higher values of such parameters, while the scatter increases at lower dust luminosities \citep{cle,pap2}. The scatter considerably reduces when selecting galaxies according to their UV absorption coefficient ($A(FUV)$). This implies that similar values of absorbed UV fraction define subsets of galaxies for which the dust-heating mechanisms are consistent. For such subsamples, the dust luminosity--SFR correlation is valid.
 \item[-] Dust mass and SFR also correlate, but more weakly than the SFR and dust luminosity.
 \item[-] The temperature of the dust in thermal equilibrium within the cold ISM correlates with the parameter $U_{min}$. A comparison with similar studies shows emissivity consistent with a value of $\beta$ = 2, but with average dust temperatures of 19.1 K, slightly higher than the measured values in the solar neighbourhood and nearby Universe. This result confirms that at higher redshifts, the dust is warmer than in the nearby Universe. As shown by \cite{mag2}, this could be due to the evolution of the metallicity combined with the increased star formation efficiency at higher redshifts \citep{tac}. However, these conclusions are strongly related to the models adopted for the dust emission and the evolution of the main sequence, and need to be interpreted with caution \citep[see also][]{dun}.
\end{itemize}

The JWST will start a new era in MIR--NIR observations, and for this reason, with the SEDs recovered from the best-fit analysis, we investigated the detection rate of our galaxies at different redshifts. At $z$ = 2, more than 60\% of the galaxies are detected with F770, F1000, F1280, F1500, and F1800. The detection decreases at higher redshifts, with only 45\% of $z$ = 8 galaxies recovered with two filters. 
Reproducing the expected sensitivity of CEERS, and classifying galaxies according to their SFR and stellar mass, we investigated the MIRI detection rate as a function of the physical properties of certain galaxies.

Fifty percent of objects with SFR $\sim$ 1 M$_\odot$yr$^{-1}$ at $z$ = 6 are detected with F770, decreasing to 20\% at $z$ = 8. For such galaxies, only 5\% of the subsample will be detected with F770 and F1000 at $z$ = 8, and only 10\% with F770, F1000, and F1280 at $z$ = 6.

In conclusion, the link between dust and star formation is very complex, and the construction of a robust data set covering a large fraction of the galaxy spectrum is mandatory for understanding the interplay between dust and stars. In this context, the MIR band plays a crucial role because it characterises both emissions and can be used to put strong constraints on the parameter space.

\begin{acknowledgements}
We warmly thank the referee for his/her constructive comments and suggestions. 
This work was supported by Fundação para a Ci\^{e}ncia e a Tecnologia (FCT) through the research grants PTDC/FIS-AST/29245/2017, UID/FIS/04434/2019, UIDB/04434/2020 and UIDP/04434/2020. C. P. acknowledges support from DL 57/2016 (P2460) from the `Departamento de F\'{i}sica, Faculdade de Ci\^{e}ncias da Universidade de Lisboa'. J.F. acknowledges financial support from the UNAM- DGAPA-PAPIIT IN111620 grant, M\'exico.

We would like to thank D. Munro for freely distributing his Yorick programming language (available at \texttt{https://github.com/LLNL/yorick}). Funding for SDSS-III has been provided by the Alfred P. Sloan Foundation, the Participating Institutions, the National Science Foundation, and the U.S. Department of Energy Office of Science. The SDSS-III web site is \texttt{http://www.sdss3.org/}. SDSS-III is managed by the Astrophysical Research Consortium for the Participating Institutions of the SDSS-III Collaboration including the University of Arizona, the Brazilian Participation Group, Brookhaven National Laboratory, Carnegie Mellon University, University of Florida, the French Participation Group, the German Participation Group, Harvard University, the Instituto de Astrofisica de Canarias, the Michigan State/Notre Dame/JINA Participation Group, Johns Hopkins University, Lawrence Berkeley National Laboratory, Max Planck Institute for Astrophysics, Max Planck Institute for Extraterrestrial Physics, New Mexico State University, New York University, Ohio State University, Pennsylvania State University, University of Portsmouth, Princeton University, the Spanish Participation Group, University of Tokyo, University of Utah, Vanderbilt University, University of Virginia, University of Washington, and Yale University. 

\end{acknowledgements}

\end{document}